\documentclass[iop]{emulateapj-rtx4}
\shortauthors{Sekanina \& Kracht}
\shorttitle{Nongravitational Motions of Dwarf Kreutz Sungrazers}
\slugcomment{Version \today }
\newcommand{\Rsun}{$R_{\mbox{\scriptsize \boldmath $\odot$}}$}
\newcommand{\Rssun}{$R_{\mbox{\boldmath $\:\!\!\scriptstyle \odot\!$}}$}

\newcommand{\gapeq}{$\;$\raisebox{0.3ex}{$>$}\hspace{-0.28cm}\raisebox{-0.75ex}{$\sim$}$\;$}

\begin{document}
\title{Strong Erosion-Driven Nongravitational Effects in Orbital
 Motions of\\the Kreutz Sungrazing System's Dwarf Comets}
\author{Zdenek Sekanina$^1$ \& Rainer Kracht$^2$}
\affil{$^1$Jet Propulsion Laboratory, California Institute of Technology,
  4800 Oak Grove Drive, Pasadena, CA 91109, U.S.A.\\
$^2$Ostlandring 53, D-25335 Elmshorn, Schleswig-Holstein, Germany}
\email{Zdenek.Sekanina@jpl.nasa.gov}

\begin{abstract}
We investigate the relationships among the angular orbital elements ---
the longitude of the ascending node $\Omega$, the inclination $i$, and the
argument of perihelion, $\omega$ --- of the Kreutz system's faint, dwarf
sungrazers observed only with the {\it SOHO/STEREO\/} coronagraphs; their
published orbits were derived using a parabolic, purely gravitational
approximation.  In a plot of $i$ against $\Omega$ the bright Kreutz sungrazers
(such as C/1843~D1, C/1882~R1, C/1963~R1, etc.) fit a curve of fixed apsidal
orientation, whereas the dwarf members are distributed along a curve that
makes with the apsidal curve an angle of 15$^\circ$.  The dwarf sungrazers'
perihelion longitude is statistically invariable, but their perihelion
latitude increases systematically with $\Omega$.  We find that this trend
can be explained by a strong erosion-driven nongravitational acceleration
normal to the orbit plane, confirmed for several test dwarf Kreutz
sungrazers by orbital solutions with nongravitational terms incorporated
directly in the equations of motion on a condition of fixed apsidal
orientation.  Proceeding in three steps, we first apply Marsden et
al.'s standard formalism, solving for the normal acceleration only,
and eventually relax additional constraints on the nongravitational law
and the acceleration's radial and transverse components.  The resulting
nongravitational accelerations on the dwarf sungrazers exceed the maximum
for catalogued comets in nearly-parabolic orbits by up to three orders of
magnitude, topping in exceptional cases the Sun's gravitational acceleration!
A mass-loss model suggests that the dwarf sungrazers' nuclei fragment
copiously and their dimensions diminish rapidly near the Sun, implying the
objects' imminent demise shortly before they reach perihelion.
\end{abstract}

\keywords{comets: general --- comets: individual (C/1843 D1, C/1880 C1,
 C/1882 R1, C/1887 B1, C/1945 X1, C/1963 R1, C/1965 S1, C/1970 K1, C/1993 A1,
 D/1993 F2, C/1998 P1, C/2001 Y4, C/2003 Q7, C/2006 J9, C/2007 X3, C/2007 X13,
 C/2008 K8, C/2008 M4, C/2008 M5, C/2009 L5, C/2011 W3) --- methods: data
 analysis{\vspace{-0.18cm}}}

\section{Introduction}
\vspace{-0.05cm}
The Kreutz system of sungrazers is by far the most prominent ensemble of
closely related comets.  Named after the German astronomer for his monumental
work on the orbital properties of its 19th century and earlier members
(Kreutz 1888, 1891, 1901), this comet system is unique.  Dynamically, the
most peculiar attribute of its members is their extremely close approach to
the Sun at perihelion, when the heliocentric distance in an overwhelming
number of cases is well below $\sim$2~{\Rsun} (1~{\Rsun} = 1~solar radius
= 0.0046548~AU) or just about 0.01~AU.  Due to perturbations, the perihelion
distance can become less than the Sun's radius, even though a {\it de facto\/}
collision with the Sun is prevented by disintegration.

All Kreutz sungrazers move about the Sun in retrograde orbits, with an
inclination in a range of \mbox{130$^\circ$--150$^\circ$}.  Their orbital
periods are, to the extent we can state, based on a few quality data
available, probably between $\sim$600 and $\sim$1000~years.  The angular
elements, pinpointing the spatial positions of the sungrazers' orbital
planes, may vary from object to object by up to at least 40$^\circ$ and
are subjected to the indirect perturbations by the planets, Jupiter in
particular.  In contrast,~the apsidal line orientation, as derived for
the bright members with reliable orbits (also referred to hereafter as
the {\it major sungrazers\/}, such as C/1843~D1, C/1882~R1, C/1963~R1,
etc.), is essentially invariable, with trivial scatter (Sec.~2).  It is
this {\it fixed position of the line of apsides\/} that Marsden (2005)
regarded as a paramount condition for defining the {\it Kreutz system's
membership\/}.  This definition reflects a common origin of all Kreutz
sungrazers and its significance is supported both by Marsden's (1967)
study of the indirect planetary perturbations (resulting in a deviation,
from one return to perihelion to the next, of typically a few tenths of
a degree in the apsidal line) and by \mbox{Sekanina}'s (2002) computation
of perturbations due to cascading fragmentation of Kreutz comets along
the orbit (similarly resulting in a difference of up to 0$^\circ\!$.02
per event for a typical separation velocity of $\sim$1 m s$^{-1}$).

No Kreutz sungrazer bright enough to observe from the ground appeared
between 1970 and 2011.  However, 19 fainter members were detected with
coronagraphs on board two satellites between 1979 and 1989 (see Marsden
2005 for a review).  This activity expanded dramatically following the
launch, in late 1995, of the {\it Solar and Heliospheric Observatory\/}
({\it SOHO\/}; Brueckner et al.\ 1995); over 2000 Kreutz sungrazers
have so far been discovered, mostly by amateur astronomers, in images
taken with the C2 and C3 coronagraphs on board {\it SOHO\/} since
January 1996.  None of these faint members of the Kreutz system, which
we hereafter refer to as {\it dwarf sungrazers\/}, survived perihelion and
none of them achieved a peak brightness greater than apparent magnitude
of about $-$0.5 (e.g., Sekanina \& Kracht 2013 and the references therein).

The beginnings of the consensus on a common origin of all Kreutz sungrazers
date back to the 19th century.  This paradigm has been strengthened by more
recent developments extensively reviewed by Marsden (2005) in a paper,
which the reader is referred to for details.  Accordingly, it is fitting
to require that the {\it invariable spatial orientation of the apsidal
line\/}, typical for the Kreutz system's bright members, {\it be exhibited
by all dwarf sungrazers\/} as well.  Strangely, this fundamental issue has
never been addressed in any detail.

Purely gravitational parabolic orbits for about 1600 dwarf Kreutz comets,
observed with the coronagraphs on board {\it SOHO\/} and the {\it Solar
Terrestrial Relations Observatory\/} ({\it STEREO\/}; Howard et al.\ 2008)
between early 1996 and mid-2010 were single-handedly computed by
Marsden.\footnote{\mbox{Most orbits are published in the {\it
Catalogue\,of\,Cometary\,Orbits\/}} \mbox{(Marsden \& Williams 2008),
$\;\!$with the rest appearing in numerous} {\it Minor Planet Circulars\/}
in the batches issued between July 2008 %(MPC 63377)
\mbox{and\,November\,2010;} %(MPC 72848-72855)
see {\tt http://www.minorplanetcenter.net/iau/
ECS/MPCArchive/MPCArchive\_TBL.html}.} This is a homogeneous set of orbits
suitable for an in-depth study of apsidal-line orientation, even though
the quality of astrometric positions measured from {\it SOHO\/} and {\it
STEREO\/} images is inferior (because of a large pixel size) compared to
the quality of ground-based observations.  One also should expect larger
uncertainties in the orbital elements of the dwarf sungrazers because of
their short orbital arcs under observation, but not any systematic trends.

Even though the orbital elements of nearly 100 additional dwarf Kreutz
sungrazers, observed with {\it SOHO\/} and {\it STEREO\/} in the second
half of 2010, have recently been published by Gray (2013), we exclude
these from our investigation to avoid mixing different orbit determination
approaches.

In this paper we employ the relationship among the angular elements ---
the longitude of the ascending node, $\Omega$; the inclination, $i$; and
the argument of perihelion, $\omega$ --- of the 1600 dwarf sungrazers to
examine in detail their compliance with the condition of fixed apsidal
orientation and to investigate the forces that affect their motions near
the Sun.  The apsidal orientations of the bright and dwarf Kreutz comets
are derived, respectively, in Secs.~2 and 3, while a perturbation analysis
of momentum changes in the orbital motions of the dwarf sungrazers is
presented in Sec.~4.  Sec.~5 compares potential interpretations of detected
variations in the apsidal orientation of the dwarf sungrazers, including
the role of erosion-driven nongravitational effects, and Sec.~6 displays
and explains the distribution of apsidal orientations from gravitational
orbital solutions.  Further refinements in a model for determining the
magnitude of an erosion-driven acceleration are introduced in Secs.~7 and 8,
where implications of the findings are also addressed.  Sec.~9 offers a
summary and conclusions, followed by our thoughts for future work in Sec.~10.

\begin{table}[ht]
\begin{center}
\vspace{0.13cm}
{\footnotesize {\bf Table 1}\\[0.1cm]
{\sc Longitude and Latitude of Perihelion for Kreutz System's\\Bright
 Members with Best Determined Orbits}\\[0.15cm]
\begin{tabular}{l@{\hspace{0.45cm}}c@{\hspace{0.3cm}}c@{\hspace{0.7cm}}l}
\hline\hline\\[-0.25cm]
         & Perihelion     & Perihelion    & \\
Comet or & longitude,     & latitude,     & Author(s) of \\
fragment & $L_\pi$(2000)  & $B_\pi$(2000) & orbital elements\\[0.08cm]
\hline\\[-0.25cm]
 & \hspace{0.19cm}$^\circ$ & \hspace{0.265cm}$^\circ$ & \\[-0.25cm]
C/1843 D1$^{\rm a}$     & 282.58 & +35.29 & Sekanina \& \\[-0.05cm]
                        &        &        & \hspace{0.1cm}Chodas (2008) \\
C/1880 C1$^{\rm b}$     & 282.38 & +35.25 & Kreutz (1901) \\
C/1882 R1-A$^{\rm c}$   & 282.94 & +35.23 & Kreutz (1891) \\
C/1882 R1-B$^{\rm c,d}$ & 282.94 & +35.23 & \hspace{0.4cm}" \\
C/1882 R1-C$^{\rm c}$   & 282.93 & +35.23 & \hspace{0.4cm}" \\
C/1882 R1-D$^{\rm c}$   & 282.94 & +35.23 & \hspace{0.4cm}" \\
C/1963 R1               & 282.65 & +35.33 & Marsden (1967) \\
C/1965 S1-A$^{\rm e}$   & 282.95 & +35.22 & \hspace{0.4cm}" \\
C/1965 S1-B             & 282.96 & +35.22 & \hspace{0.4cm}" \\
C/1970 K1               & 282.95 & +35.07 & Marsden (1970) \\
C/2011 W3               & 282.98 & +35.09 & Sekanina \& \\[-0.05cm]
                        &        & & \hspace{0.1cm}Chodas (2012) \\[0.05cm]
\hline\\[-0.25cm]
Average$^{\rm f}$       & 282.81 & +35.22 & \\[-0.05cm]
                        & $\,\pm$0.21 & $\:\:\pm$0.08 & \\[0.05cm]
\hline\\[-0.25cm]
\multicolumn{4}{l}{\parbox{8.2cm}{$^{\rm a}$\,{\scriptsize Solution
II.}}}\\[-0.1cm]
\multicolumn{4}{l}{\parbox{8.2cm}{$^{\rm b}$\,{\scriptsize Solution
 B.}}}\\[-0.05cm]
\multicolumn{4}{l}{\parbox{8.2cm}{$^{\rm c}$\,{\scriptsize Solution
 III.}}}\\[-0.04cm]
\multicolumn{4}{l}{\parbox{8.2cm}{$^{\rm d}$\,{\scriptsize Nonrelativistic
 solution; {\vspace*{-0.05cm}}a relativistic solution by Hufnagel (1919)
 differs at most in the fifth decimal.}}}\\[0.18cm]
\multicolumn{4}{l}{\parbox{8.2cm}{$^{\rm e}$\,{\scriptsize Nonrelativistic
 solution; {\vspace*{-0.05cm}}a relativistic solution by Marsden (1967)
 differs at most in the fourth decimal.}}}\\[0.09cm]
\multicolumn{4}{l}{\parbox{8.2cm}{$^{\rm f}$\,{\scriptsize Companions
A, C, and D of comet C/1882 R1 and{\vspace*{-0.05cm}} companion B of
comet C/1965 S1 are given half weight.}}}\\[0cm]
\end{tabular}}
\end{center}
\end{table}

\section{Line of Reference Apsidal Orientation}
High-quality sets of orbital elements are available only for seven bright
Kreutz sungrazers seen since the 1840s.  All of them were more or less
spectacular objects visible with the unaided eye and observed
astrometrically over mostly extended periods of time from the ground,
and their orbits have the lines of apsides nearly perfectly aligned.
Table~1, which lists their perihelion longitude $L_\pi$ and latitude
$B_\pi$, shows that a mean perihelion point is described by
\begin{eqnarray}
\langle L_\pi \rangle & = & 282^\circ\!.8 \pm 0^\circ\!.2, \nonumber \\[-0.25cm]
                      &   & \\[-0.25cm]
\langle B_\pi \rangle & = & +35^\circ\!.2 \pm 0^\circ\!.1 \nonumber
\end{eqnarray}
(Equinox J2000.0).  The maximum deviations from the adopted mean values
amount to 0$^\circ\!$.4 in the longitude and less than 0$^\circ\!$.2 in
the latitude.  The curve that in plots of $i$ against $\Omega$ and of
$\omega$ against $\Omega$ fits the coordinates (1) is hereafter called the
{\it line of reference apsidal orientation\/} and is expressed, respectively
in the two plots, by
\begin{eqnarray}
\cot i & = & \cot \langle B_\pi \rangle \sin ( \langle L_\pi
 \rangle - \Omega ), \nonumber \\[-0.23cm]
         &    &  \\[-0.23cm]
\cos \omega & = & \cos \langle B_\pi \rangle \cos ( \langle
L_\pi \rangle - \Omega ). \nonumber
\end{eqnarray}
Excluded from Table 1 are the headless object C/1887 B1 and C/1945 X1,
whose orbit is known less accurately:\ the solution that Marsden (1967)
considered the best differs from the mean by +0$^\circ\!$.7 in $L_\pi$
and +0$^\circ\!$.8 in $B_\pi$.

\section{Relationships Among the Angular Elements of the Dwarf Kreutz
Sungrazers}
\begin{figure}[b]
\vspace*{0.4cm}
\hspace*{-0.69cm}
\centerline{
\scalebox{0.535}{
\includegraphics{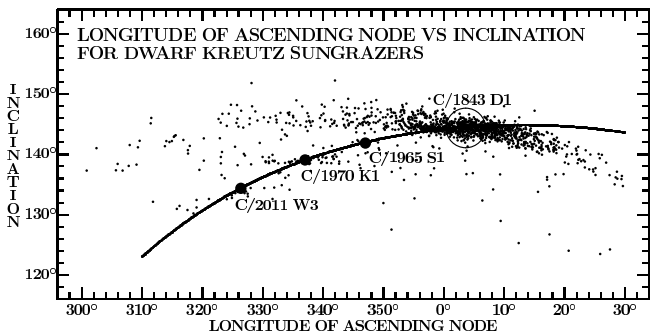}}} % from f1_SOHO-rev.tex
\vspace*{-9.2cm}
\caption{Plot of the longitude of the ascending node against the
orbit inclination for 1565 dwarf Kreutz sungrazers from the
period 1996 January to 2010 June.  Also plotted are four major
members of the Kreutz system:\ C/1843 D1 (in the center~of the
oversized open circle), C/1965 S1, C/1970 K1, and C/2011~W3.  The
solid curve is the line of reference apsidal orientation, described
by \mbox{$\langle L_\pi \rangle = 282^\circ\!.8$} and \mbox{$\langle
B_\pi \rangle = +35^\circ\!.2$} (eq.\ J2000).}
\end{figure}

A near-perfect alignment of the lines of apsides is unfortunately not
what one finds when examining the relationship between the longitude of
the ascending node and the inclination of the dwarf Kreutz sungrazers,
as is readily apparent from Figure 1.  An overwhelming majority of these
objects in the plot is distributed along an arc that passes through the
location occupied by C/1843 D1 and makes a sizable angle with the curve of
reference apsidal orientation.  Smaller numbers of the dwarf Kreutz sungrazers
are also distributed along parallel arcs passing through the locations of
C/1970~K1 and C/2011~W3, but much less so through C/1965~S1.

Similarly, in Figure 2 we plot the longitude of the ascending node against
the argument of perihelion for the same set of the dwarf Kreutz sungrazers.
This time, the dwarf and the major members appear to be distributed along
an essentially common curve.

\begin{figure}[b]
\vspace*{0.4cm}
\hspace*{-0.69cm}
\centerline{
\scalebox{0.535}{
\includegraphics{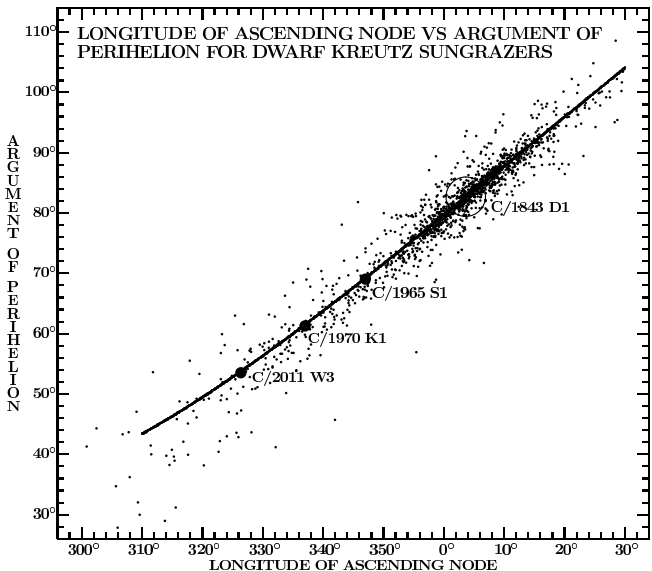}}} % from f2_SOHO-rev.tex
\vspace*{-5.45cm}
\caption{Plot of the longitude of the ascending node against the
argument of perihelion for 1565 dwarf Kreutz sungrazers from
the period 1996 January to 2010 June.  Also plotted are the four
major members of the Kreutz system, as in Figure 1.  The solid
curve is again the line of reference apsidal orientation (see the
caption to Figure 1).}
\end{figure}

\begin{figure}[b]
\vspace*{0.45cm}
\hspace*{-1.3cm}
\centerline{
\scalebox{0.545}{
\includegraphics{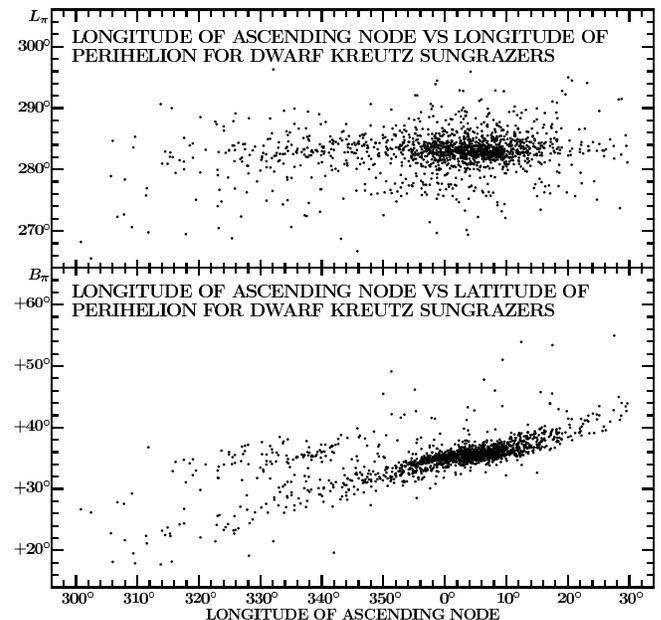}}} % from f3_SOHO-rev.tex
\vspace*{-3.9cm}
\caption{Plot of the longitude $L_\pi$ (at the top) and latitude $B_\pi$
of perihelion for 1565 dwarf Kreutz sungrazers from the period
1996 January to 2010 June.  While the longitude stays constant
over an interval  of nearly 90$^\circ$ in the nodal longitude, the
latitude increases systematically, for both the thickly populated
branch and the other two branches.}
\end{figure}

\begin{table*}
\vspace{0.1cm}
\begin{center}
{\footnotesize {\bf Table 2}\\[0.1cm]
{\sc Mean Orbital Elements $\omega$, $\Omega$, $i$, $q$, and Longitude and
 Latitude of Perihelion (Eq.\ 2000):\\Branch of Dwarf Kreutz Sungrazers
 Fitting Orbit of Comet C/1843 D1.}\\[0.1cm]
\begin{tabular}{c@{\hspace{0.25cm}}c@{\hspace{0.3cm}}c@{\hspace{0.5cm}}c@{\hspace{0.5cm}}c@{\hspace{0.5cm}}c@{\hspace{0.65cm}}c@{\hspace{0.45cm}}c}
\hline\hline\\[-0.25cm]
Interval of & Number
 & \multicolumn{4}{@{\hspace{-0.45cm}}c}{Mean value of element$^{\rm a}$}
 & \multicolumn{2}{@{\hspace{-0.01cm}}c}{Line of apsides} \\[-0.06cm]
ascending & of sun-
 & \multicolumn{4}{@{\hspace{-0.45cm}}c}{\rule[0.6ex]{7.3cm}{0.4pt}}
 & \multicolumn{2}{@{\hspace{-0.01cm}}c}{\rule[0.6ex]{2cm}{0.4pt}} \\[-0.06cm]
nodes, $\Omega$ & grazers & $\omega$ & $\Omega$ & $i$ & $q$ ($R_{\mbox{\boldmath
 $\scriptstyle \odot$}}$)$^{\rm b}$ & $L_\pi$ & $B_\pi$ \\[0.05cm]
\hline\\[-0.22cm]
300$^\circ\!$--\,310\rlap{$^\circ$} & $\;\:$5
 & 32$^\circ\!\!$.1$\,\pm\,$3$^\circ\!\!$.4
 & 307$^\circ\!\!$.7$\,\pm\,$1$^\circ\!\!$.8
 & 140$^\circ\!\!$.0$\,\pm\,$2$^\circ\!\!$.1
 & 1.40$\:\pm\:$0.31 & 282$^\circ\!\!$.1 & +19$^\circ\!\!$.9 \\
310\,--\,320 & $\;\:$7 & 39.8$\:\pm\:$1.2 & 315.5$\:\pm\:$1.3
 & 142.0$\:\pm\:$1.0 & 1.28$\:\pm\:$0.18 & 282.2 & +23.2 \\
320\,--\,330 & 15 & 47.8$\:\pm\:$3.5 & 325.7$\:\pm\:$2.0
 & 145.9$\:\pm\:$1.6 & 1.35$\:\pm\:$0.27 & 283.3 & +24.5 \\
330\,--\,335 & 13 & 56.4$\:\pm\:$2.5 & 332.8$\:\pm\:$1.4
 & 145.2$\:\pm\:$1.6 & 1.35$\:\pm\:$0.34 & 281.8 & +28.4 \\
335\,--\,340 & 14 & 59.0$\:\pm\:$2.4 & 337.2$\:\pm\:$1.4
 & 145.6$\:\pm\:$1.4 & 1.27$\:\pm\:$0.22 & 283.3 & +28.9 \\
340\,--\,345 & 28 & 63.6$\:\pm\:$1.9 & 342.9$\:\pm\:$1.5
 & 146.0$\:\pm\:$1.3 & 1.20$\:\pm\:$0.20 & 283.8 & +30.1 \\
345\,--\,350 & 37 & 68.5$\:\pm\:$2.3 & 347.3$\:\pm\:$1.3
 & 145.7$\:\pm\:$1.3 & 1.21$\:\pm\:$0.20 & 282.8 & +31.6 \\
350\,--\,355 & 65 & 73.3$\:\pm\:$1.8 & 353.0$\:\pm\:$1.4
 & 145.3$\:\pm\:$1.1 & 1.18$\:\pm\:$0.20 & 283.0 & +33.0 \\
355\,--\,$\;\;\:\:$0 & \llap{1}78 & 77.9$\:\pm\:$1.4 & 358.1$\:\pm\:$1.3
 & 144.6$\:\pm\:$0.9 & 1.12$\:\pm\:$0.13 & 282.8 & +34.5 \\
$\;\;\:\:$0\,--\,$\;\;\:\:$5 & \llap{2}50 & 81.6$\:\pm\:$1.4
 & $\;\;\:\:$2.6$\:\pm\:$1.4 & 144.4$\:\pm\:$0.7 & 1.12$\:\pm\:$0.13
 & 282.9 & +35.2 \\
$\;\;\:\:$5\,--\,$\;\:$10 & \llap{2}65 & 85.5$\:\pm\:$1.2
 & $\;\;\:\:$7.3$\:\pm\:$1.4 & 144.3$\:\pm\:$0.6 & 1.15$\:\pm\:$0.16
 & 282.8 & +35.6 \\
$\;\:$10\,--\,$\;\:$15 & \llap{1}30 & 89.1$\:\pm\:$1.4 & $\;\:$12.2$\:\pm\:$1.3
 & 143.0$\:\pm\:$1.2 & 1.19$\:\pm\:$0.19 & 283.4 & +37.0 \\
$\;\:$15\,--\,$\;\:$20 & 50 & 92.3$\:\pm\:$1.6 & $\;\:$16.8$\:\pm\:$1.3
 & 141.8$\:\pm\:$1.1 & 1.17$\:\pm\:$0.19 & 283.8 & +38.1 \\
$\;\:$20\,--\,$\;\:$30 & $\;\:$7 & \llap{1}02.9$\:\pm\:$3.0
 & $\;\:$28.3$\:\pm\:$1.7 & 136.2$\:\pm\:$0.8 & 1.18$\:\pm\:$0.20
 & 280.7 & +42.4 \\
$\;\:$30\,--\,$\;\:$40 & 12 & \llap{1}05.0$\:\pm\:$3.8 & $\;\:$34.7$\:\pm\:$2.4
 & 133.7$\:\pm\:$1.7 & 1.27$\:\pm\:$0.29 & 283.5 & +44.3 \\[0.05cm]
\hline\\[-0.29cm]
\multicolumn{8}{l}{\parbox{10cm}{$^{\rm a}$\,{\scriptsize Samples also
subjected to some limitation of intervals in $\omega$ and $i$.}}}\\[-0.07cm]
\multicolumn{8}{l}{\parbox{10cm}{$^{\rm b}$\,{\scriptsize The solar radius,
$R\!_{\mbox{\boldmath $\scriptscriptstyle \odot$}}$, is equivalent to
0.0046548 AU.}}}
\end{tabular}}
\end{center}
\end{table*}

We thus find that for some reason the orbital distribution of the dwarf
sungrazers differs significantly from the distribution of the major
members of the Kreutz system in the plot of the longitude of the
ascending node $\Omega$ against the inclination $i$ but not against the
argument of perihelion $\omega$.  The roots of the inconsistence between
the behavior of the major and the dwarf sungrazers in Figure 1 are
apparent from Figure 3 in a plot of the orientation of the apsidal
line of the dwarf comets' against the position of their nodal line, that
is, of $L_\pi$ and $B_\pi$ against $\Omega$.  Except for the scatter,
the longitude $L_\pi$ is seen to be essentially constant over a span of
nearly 90$^\circ$ in $\Omega$.  In sharp contrast, the latitude $B_\pi$
{\it increases systematically with $\Omega$\/}.  In fact, this trend
has an effect on the plot in Figure 2 as well, but --- as explained in
Sec.~4, only a minor one.  In addition to the thickly populated branch,
which fits the position of comet C/1843 D1, we recognize in Figure~3,
just as in Figure~1, the two thinly populated branches that pass,
respectively, through the positions of C/1970~K1 and C/2011~W3.

By dividing the entire range of the longitudes of the ascending node into
a number of intervals, the same effect is displayed in Table 2 for the
thickly populated branch of the dwarf Kreutz sungrazers and in Table 3
for the other two branches.  Averaging the values of $L_\pi$ in the
penultimate column of Table 2 yields 282$^\circ\!$.8\,$\pm$\,0$^\circ\!$.8,
in perfect agreement with the adopted value of $\langle L_\pi \rangle$
based on the orbits of the major Kreutz-system members, while the values
of $B_\pi$ increase systematically by nearly 25$^\circ$!{\hspace{0.3cm}}

In order to separate the three branches from one another when calculating
the mean values of the elements within each interval of $\Omega$, it was
necessary also to restrict the range of corresponding inclinations.  Once
this was done, it was advisable also to limit $\omega$ to a certain range
of values, even though Figure 2 provides no obvious clue as to where to
draw the boundaries.  In practice we followed a simple rule:\ we first
incorporated all dwarf Kreutz sungrazers picked up by the computer code
in the given range of $\Omega$; next we inspected the inclinations of
all these entries and eliminated those (if any) judged to be clearly
out of acceptable bounds; and finally we checked the arguments of
perihelion of the remaining sungrazers and again removed those that
appeared to be out of bounds to get the final set for each interval of
$\Omega$.  Admittedly, this sort of approach is always somewhat arbitrary,
but only at the periphery of each set.

While the two sets in Table 3 substantially support the results based on
the data presented in Table 2, there are some differences.  These are
likely to be due in part to the fact that the sets in Table 3 are much
smaller and the errors often (though not always) larger.  The average
values of the angular elements are in both sets again quite close to the
values for the respective major comets, but the perihelion distance of the
dwarf Kreutz sungrazers in the branch fitting C/1970 K1 is by more than
1$\sigma$ lesser than the comet's distance and the opposite is true for
the sungrazers in the branch fitting C/2011 W3.  Because of the large
uncertainties in the perihelion distance (which may be greatly
underestimated by the formal errors calculated from the scatter among
the individual objects), these discrepancies may not be significant.

\begin{table*}
\vspace{0.1cm}
\begin{center}
{\footnotesize {\bf Table 3}\\[0.1cm]
{\sc Mean Orbital Elements $\omega$, $\Omega$, $i$, $q$, and Longitude and
Latitude of Perihelion (Eq.\ 2000):\\Branches of Dwarf Kreutz Sungrazers
Fitting Orbits of C/1970 K1 or C/2011 W3.}\\[0.1cm]
\begin{tabular}{c@{\hspace{0.25cm}}c@{\hspace{0.3cm}}c@{\hspace{0.5cm}}c@{\hspace{0.5cm}}c@{\hspace{0.5cm}}c@{\hspace{0.65cm}}c@{\hspace{0.45cm}}c}
\hline\hline\\[-0.25cm]
Interval of & Number
 & \multicolumn{4}{@{\hspace{-0.45cm}}c}{Mean value of element$^{\rm a}$}
 & \multicolumn{2}{@{\hspace{-0.01cm}}c}{Line of apsides} \\[-0.06cm]
ascending & of sun-
 & \multicolumn{4}{@{\hspace{-0.45cm}}c}{\rule[0.6ex]{7.3cm}{0.4pt}}
 & \multicolumn{2}{@{\hspace{-0.01cm}}c}{\rule[0.6ex]{2cm}{0.4pt}} \\[-0.06cm]
nodes, $\Omega$ & grazers & $\omega$ & $\Omega$ & $i$ & $q$ ({\Rssun})
 & $L_\pi$ & $B_\pi$ \\[0.05cm]
\hline\\[-0.12cm]
 & & & \rlap{\hspace{-0.75cm}\sf Branch Fitting C/1970 K1} & & & & \\[0.08cm]
320$^\circ\!$--\,330\rlap{$^\circ$} & $\;\:$6
 & 53$^\circ\!\!$.0$\,\pm\,$3$^\circ\!\!$.4
 & 324$^\circ\!\!$.1$\,\pm\,$2$^\circ\!\!$.9
 & 138$^\circ\!\!$.5$\,\pm\,$1$^\circ\!\!$.2
 & 1.49$\:\pm\;$0.20 & 279$^\circ\!\!$.3 & +31$^\circ\!\!$.9 \\
330\,--\,335 & 20 & 58.5$\:\pm\:$2.8 & 332.3$\:\pm\:$1.6 & 138.4$\:\pm\:$1.0
 & 1.59$\:\pm\:$0.26 & 281.7 & +34.5 \\
335\,--\,340 & 27 & 62.0$\:\pm\:$3.2 & 337.7$\:\pm\:$1.4 & 139.0$\:\pm\:$0.6
 & 1.63$\:\pm\:$0.22 & 282.9 & +35.4 \\
340\,--\,345 & 18 & 66.6$\:\pm\:$3.6 & 341.8$\:\pm\:$1.3 & 139.6$\:\pm\:$1.0
 & 1.51$\:\pm\:$0.29 & 281.4 & +36.5 \\
345\,--\,355 & 14 & 72.0$\:\pm\:$3.9 & 348.5$\:\pm\:$2.3 & 140.2$\:\pm\:$1.8
 & 1.44$\:\pm\:$0.29 & 281.4 & +37.5 \\[0.11cm]
 & & & \rlap{\hspace{-0.75cm}\sf Branch Fitting C/2011 W3} & & & & \\[0.08cm]
315$^\circ\!$--\,320\rlap{$^\circ$} & $\;\:$6
 & 47$^\circ\!\!$.5$\,\pm\,$1$^\circ\!\!$.5
 & 317$^\circ\!\!$.7$\,\pm\,$1$^\circ\!\!$.4
 & 131$^\circ\!\!$.1$\,\pm\,$0$^\circ\!\!$.7
 & 1.37$\:\pm\:$0.28 & 282$^\circ\!\!$.0 & +33$^\circ\!\!$.8 \\
320\,--\,325 & $\;\:$4 & 50.9$\:\pm\:$2.1 & 323.7$\:\pm\:$1.8
 & 132.7$\:\pm\:$0.3 & 1.45$\:\pm\:$0.31 & 283.9 & +34.8 \\
325\,--\,330 & $\;\:$7 & 54.5$\:\pm\:$1.7 & 326.9$\:\pm\:$0.9
 & 133.6$\:\pm\:$0.4 & 1.48$\:\pm\:$0.18 & 282.9 & +36.2 \\
330\,--\,335 & $\;\:$9 & 58.5$\:\pm\:$1.8 & 332.6$\:\pm\:$1.6
 & 135.5$\:\pm\:$1.2 & 1.48$\:\pm\:$0.33 & 283.3 & +36.7 \\[0.06cm]

\hline\\[-0.3cm]
\multicolumn{8}{l}{\parbox{10cm}{$^{\rm a}$\,{\scriptsize Samples also
 subjected to some limitation of intervals in $\omega$ and
 $i$.}}}\\[-0.15cm]
\end{tabular}}
\end{center}
\end{table*}

Tables 2 and 3 and Figure 3 show some differences between the three
branches of the dwarf Kreutz sungrazers in the rate of $\Delta
B_\pi/\Delta \Omega$.  On the average, the rate is \mbox{$\langle
\Delta B_\pi/\Delta \Omega \rangle = +0.28$} for the set in Table~2,
but +0.23 and +0.19, respectively, for the two sets in Table~3.  The
validity of this comparison may be questioned because the overall ranges
of the longitude of the ascending node for the three sets are very
different.  However, in the intervals of $\Omega$ covered by the two
sets in Table~3, the thickly populated branch offers for $\langle \Delta
B_\pi/\Delta \Omega \rangle$ even higher values than is the average,
+0.33 and +0.30, respectively.  Thus, this effect appears to be genuine.
Pure\-ly empirically, one can argue that in a clockwise rotation in
Figure 1, starting from the line of reference apsidal orientation toward
the thickly populated branch of sungrazers, the rate $\Delta B_\pi/\Delta
\Omega$ keeps increasing systematically from zero to a maximum.  Because
the two sparsely populated branches are located in between these two, their
rates are intermediate between zero and the maximum.  In the lower panel
of Figure 3 one can imagine a line parallel to the axis of abscissae
having an ordinate of +35$^\circ\!$.2 and populated by the major
sungrazers.  Thus, {\it each of the three branches, along which the
overwhelming majority of the dwarf Kreutz sungrazers is distributed
in Figure~1, will coincide with the line of reference apsidal orientation
if they are rotated counterclockwise around the respective major sungrazer's
(C/1843~D1, C/1970~K1, or C/2011~W3) position by the same angle, about
15\,$^\circ\!$}.  This is equivalent to a rotation of the dwarf sungrazers'
orbital planes by an $\Omega$-dependent amount, needed to eliminate the
systematic trends in the latitude of perihelion $B_\pi$.

To show how well this simple-minded rotation works, Figure 4 displays the
same plot of the dwarf Kreutz sungrazers as Figure 1, except that
the line of reference apsidal orientation is now rotated clockwise by
15$^\circ$ and forced to pass, respectively, through the location in the
plot of comet C/1843~D1 (top branch; curve {\it A\/}), through the location
of C/1970~K1 (middle branch; curve {\it B\/}), and through a point close
to the location of C/2011~W3 (lowest branch; curve~{\it C\/}).

\begin{figure}[b]
\vspace{0.3cm}
\hspace{-0.64cm}
\centerline{
\scalebox{0.525}{
\includegraphics{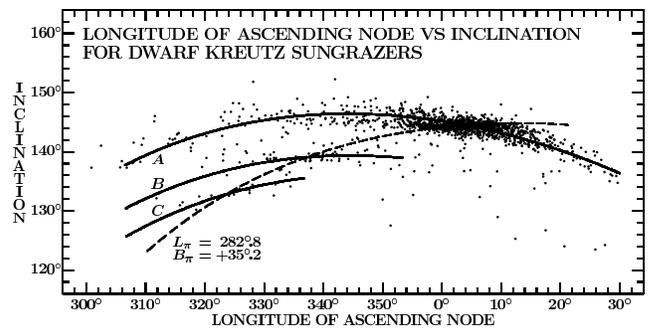}}} % from f4_SOHOrev.tex
\vspace{-9.15cm}
\caption{Plot of the longitude of the ascending node against the
orbit inclination, as in Figure 1, with the three branches of the
dwarf Kreutz sungrazers (dots) being compared not only to the
line of reference apsidal orientation (thin dashed curve) but also
to its {\it rotated\/} versions, marked {\it A}, {\it B}, and {\it C},
that pass, respectively, through the positions of three major Kreutz
sungrazers:\ C/1843~D1, C/1970~K1, and, approximately, C/2011~W3.  The
angle of rotation, 15$^\circ$, is the same for all three populations of
the dwarf sungrazers.  Thus, the match, which is very satisfactory, is
{\it not\/} coming from, and does {\it not\/} represent, least-squares
solutions.}
\end{figure}

To summarize, the systematic trend in the perihelion latitude of the
dwarf sungrazers appears to be linked to some effects that the bright
members of the Kreutz system manage to avoid.  And because the dwarf
sungrazers could not, as individual bodies, survive longer than one
revolution about the Sun (otherwise they should be seen to recede from
the Sun after perihelion), the effects in Figures~1 and 3 and in Tables~2
and 3 should be a product of the dwarf sungrazers' evolution in the course
of this revolution about the Sun.

Tables 2 and 3 also present the averaged perihelion distance $q$ of
the dwarf sungrazers in each interval of the longitude of the
ascending node.  We find that in the thickly populated branch the
minimum nominal distance of $\sim$1.1 {\Rsun} is reached near
\mbox{$\Omega \simeq 0^\circ$} but that the minima are closer to
1.4--1.5 {\Rsun} in the two thinly populated branches.  Because of
the uncertainties involved and also because the orbits of some of
these objects were derived with the perihelion distance being forced
by Marsden to particular values (Sekanina 2002; see also Sec.~9), one
should not attach much significance to these variations.

A more interesting result is found from inspection of the entries
for the interval of \mbox{$0^\circ\! < \Omega < 5^\circ$} in Table~2.
All five tabulated angles equal, within the limits of error, the
respective angular elements of comet C/1843~D1 (Table~1 and Sekanina
\& Chodas 2008).  This match~strongly suggests a close relationship
between this comet and the most populated branch of the dwarf Kreutz
sungrazers despite the 150+ years spanned between them.

The contrast between the plot of $i$ against $\Omega$ (Figure 1), in which
the dwarf Kreutz sungrazers behave differently from the major members,
and the plot of $\omega$ against $\Omega$, in which the two categories of
objects almost overlap, is illustrated best by the calculated values of the
elements at the ends of the $\Omega$ range in Table 2.  At \mbox{$\Omega =
307^\circ\!.7$} the difference between the major and the dwarf members
is nearly $-$20$^\circ$ in $i$ but only +10$^\circ$ in $\omega$, while at
\mbox{$\Omega = 34^\circ\!.7$} it is +9$^\circ$ in $i$ but about $-$3$^\circ$
in $\omega$.  We next address~these discrepancies, as well as the differences
in the behavior of $L_\pi$ and $B_\pi$ of the dwarf Kreutz sungrazers in
Figure 3 and Tables 2 and 3, in terms of perturbations caused by momentum
changes in the dwarf sungrazers' motions.

\section{Perturbations Due to Momentum Changes\\in the Orbital Motions of
the\\Dwarf Kreutz Sungrazers}

The enormous population of the dwarf Kreutz sungrazers could have arisen
on the timescale of a single revolution about the Sun only by cascading
fragmentation of larger objects in the Kreutz system, which are likely
to have included some very massive members.  One can hypothesize that, as
a consequence of numerous fragmentation events, the orbits of the fragments
have gradually been modified compared to the initial parent's orbit, and
that their observed distribution reflects a cumulative effect of these
modifications.  If so, the question then arises about the nature of these
effects that (a)~show up in the plot of $\Omega$ against $i$, but not (at
least nowhere as prominently) in the plot of $\Omega$ against $\omega$, and
(b)~manifest themselves as systematic variations in the perihelion latitude
$B_\pi$ but not in the longitude $L_\pi$.

Aiming to examine the nature of the effect in Figures 1 and 3 and in Tables 2
and 3, we begin with the relations between (a)~the instantaneous rate of
change, at time $t$, of the angular orbital elements --- $d\omega/dt$,
$d\Omega/dt$, $di/dt$ --- and (b)~the momentum acquired during a breakup,
expressed in terms of the components of an acceleration $j(t)$
imparted to a fragment relative to the parent body in the three cardinal
directions defined by the heliocentric orbit of the parent in the radial
(away from the Sun), R, transverse, T, and normal, N, directions of the
right-handed RTN coordinate system.  These relations can be written thus
(e.g., Danby 1988):
\begin{equation}
\left( \!\! \begin{array}{c}
 {\displaystyle \frac{d\omega}{dt}}\\[0.33cm]
 {\displaystyle \frac{d\Omega}{dt}}\\[0.3cm]
 {\displaystyle \frac{di}{dt}}
 \end{array} \!\! \right) = \left( \! \begin{array}{ccc}
 X_{\rm R} & X_{\rm T} & X_{\rm N} \\
 Y_{\rm R} & Y_{\rm T} & Y_{\rm N} \\
 Z_{\rm R} & Z_{\rm T} & Z_{\rm N}
 \end{array} \!\! \right) \! \cdot \! \left(\! \begin{array}{c}
 j_{\rm R} \\ j_{\rm T} \\ j_{\rm N}
 \end{array} \!\! \right) \!\! ,
\end{equation}
where
\begin{equation}
\begin{array}{l}
X_{\rm R} = {\displaystyle \frac{\chi}{e^2}} \, (1 - \psi), \\[0.3cm]
X_{\rm T} = {\displaystyle \frac{\chi}{e \psi}} \, (1+\psi) \, \sin u, \\[0.3cm]
X_{\rm N} = -{\displaystyle \frac{\chi}{\psi}} \, \cot i \,
 \sin(\omega + u), \\[0.35cm]
Y_{\rm N} = {\displaystyle \frac{\chi}{\psi} \;
 \frac{\sin(\omega+u)}{\sin i}}, \\[0.35cm]
Z_{\rm N} = {\displaystyle \frac{\chi}{\psi}} \, \cos(\omega + u),
\end{array}
\end{equation}
and
\begin{equation}
Y_{\rm R} = Y_{\rm T} = Z_{\rm R} = Z_{\rm T} = 0,
\end{equation}
with $\chi = \sqrt{p}/k$, $\psi = p/r$, $k$ being the Gaussian gravitational
constant, $p$ the parameter of the orbit, $p = q\,(1 + e)$, $q$ the perihelion
distance, $e$ the eccentricity of the orbit, and \mbox{$r = r(t)$} and
\mbox{$u = u(t)$}, respectively, the heliocentric distance and the true
anomaly at time $t$.

Since the fragmentation process is assumed to consist of discrete, short-term
events, of a duration \mbox{$\Delta t \rightarrow 0$} and our interest is only
in their integrated outcome, we replace the derivative $d\omega/dt$ on the
left-hand side of Eq.\,(3) with
\begin{equation}
\Delta \omega = \int_{(\Delta t \rightarrow 0)} \!\! {\displaystyle
 \frac{d\omega(t)}{dt}} \; dt
\end{equation}
and similarly $\Delta \Omega$ and $\Delta i$; and, on the right-hand
side,~we introduce the components of the separation velocity,
\begin{equation}
V_{\rm R} = \int_{(\Delta t \rightarrow 0)} \! j_{\rm R}(t) \; dt,
\end{equation}
and similarly $V_{\rm T}$ and $V_{\rm N}$.  Since the events are brief,
the heliocentric distance $r$ and the true anomaly $u$ are constants (i.e.,
\mbox{$\Delta r \rightarrow 0$} and \mbox{$\Delta u \rightarrow 0$}).
Equation (3) now becomes
\begin{equation}
\left( \!\! \begin{array}{c}
\Delta \omega \\ \Delta \Omega \\ \Delta i
\end{array} \!\! \right) = \left( \!\!
\begin{array}{ccc}
X_{\rm R} & X_{\rm T} & X_{\rm N} \\
0 & 0 & Y_{\rm N} \\
0 & 0 & Z_{\rm N}
\end{array} 
\!\! \right) \! \cdot \! \left( \!\!
\begin{array}{c}
V_{\rm R} \\ V_{\rm T} \\ V_{\rm N}
\end{array}
\!\! \right) \!\! .
\end{equation}
Increments $\Delta \omega$, $\Delta \Omega$, and $\Delta i$ affect
increments $\Delta L_\pi$ and $\Delta B_\pi$ in the direction to
perihelion, which are equal
to
\begin{equation}
\begin{array}{c}
\Delta L_\pi = \Delta \Omega + {\displaystyle \frac{\cos i}{1 - \sin^2
 \omega \sin^2 i}} \left( \Delta \omega \! - \! \frac{1}{2} \Delta i
 \, \sin 2\omega \, \tan i \right) \! , \\[0.4cm]
\Delta B_\pi = {\displaystyle \frac{\cos \omega \sin i}{\sqrt{1 - \sin^2
 \omega \sin^2 i}}} \left( \Delta \omega + \Delta i \, \tan \omega \, \cot i
 \right) \! .
\end{array}
\end{equation}
To the extent that Figure 3 implies that, on the average, \mbox{$\Delta
L_\pi = 0$}, the following constraint on $V_{\rm R}$, $V_{\rm T}$, and
$V_{\rm N}$ applies after inserting for $\Delta \omega$, $\Delta \Omega$,
and $\Delta i$ from (8) and (4) into the first equation of (9):
\begin{equation}
X_{\rm R} V_{\rm R} \! + \! X_{\rm T} V_{\rm T} = - \Psi \cos \omega \,
 \tan i \, \sin u \cdot V_{\rm N},
\end{equation}
where
\begin{equation}
\Psi = \frac{\chi}{\psi} = \frac{r}{k \sqrt{p}}.
\end{equation}
The second equation of (9) can similarly be rewritten as
\begin{equation}
\begin{array}{llll}
\Delta B_\pi & \!\!\! = \!\!\! & {\displaystyle \frac{\cos \omega
 \sin i}{\sqrt{1 - \sin^2 \omega \, \sin^2 i}}} \\[0.45cm]
 & & \times \left( X_{\rm R} V_{\rm R} \!+\! X_{\rm T} V_{\rm T} - \Psi
 \sec \omega \, \cot i \, \sin u \cdot V_{\rm N} \right) \! .
\end{array}
\end{equation}
Inserting for \mbox{$X_{\rm R} V_{\rm R} \!+\! X_{\rm T} V_{\rm T}$} from
(10) into (12) we get the final expression for $\Delta B_\pi$:
\begin{equation}
\Delta B_\pi = -\Psi {\displaystyle \frac{\cos i \, \sin u}{\sqrt{1 -
 \sin^2 \omega \, \sin^2 i}}} \left( 1 \!+ \cos^2 \omega \, \tan^2 i
 \right) \cdot V_{\rm N}.
\end{equation}
This equation says that as long as the perihelion longitude $L_\pi$ of the
dwarf Kreutz sungrazers in Figure~1 is statistically independent of the
three angular orbital elements, the {\it systematic increment in the
perihelion latitude is only a function of a momentum change in the
direction normal to the orbital plane\/}.

On condition (10), the increments in all three angular elements depend also
only on the normal component,
\begin{equation}
\begin{array}{c}
\Delta \omega = -\Psi \cot i \! \left[ \sin (\omega \!+\! u) + \cos \omega
 \tan^2 i \sin u \right] V_{\rm N}, \\[0.05cm]
\Delta \Omega = \Psi \: {\displaystyle \frac{\sin (\omega \!+\! u)}{\sin i}}
 \, V_{\rm N}, \\[0.2cm]
\Delta i = \Psi \, \cos (\omega \!+\! u) \: V_{\rm N}.
\end{array}
\end{equation}
The expressions for the ratio of increments of the angles plotted in Figures
1 and 2 we obtain, respectively,
\begin{equation}
\frac{\Delta i}{\Delta \Omega} = \sin i \, \cot (\omega \!+\! u) \approx
 \sin i \cot \omega,
\end{equation}
and
\begin{equation}
\frac{\Delta \omega}{\Delta \Omega} = -\cos i \! \left[ 1 + \frac{\cos \omega
 \, \tan^2 i \, \sin u}{\sin (\omega \!+\! u)} \right] \approx -\cos i,
\end{equation}
where the approximations reflect the fact that the true anomaly
\mbox{$u \rightarrow \pm 180^\circ$} everywhere in the orbit except very
close to perihelion.  Equations (15) and (16) can also be derived directly
from Eqs.\,(2).  Now, there is one major difference between Eq.\,(15) and
Eq.\,(16).  As a function of merely the inclination, which varies within
fairly tight limits, mostly between $\sim$130$^\circ$ and $\sim$150$^\circ$
(Figure~1), the ratio $\Delta \omega/\Delta \Omega$ varies between +0.64
and +0.87, which explains nearly linear relationship between the two
elements with an average slope of +0.75 in Figure~2.  By contrast, the ratio
$\Delta i/\Delta \Omega$ depends on both the inclination and the argument of
perihelion, which varies widely, from $\sim$30$^\circ$ to $\sim$110$^\circ$
(Figure 2).  Hence, $\Delta i/\Delta \Omega$ can be of either sign and the
curve of reference apsidal orientation in the plot of $i = i(\Omega)$ reaches
a maximum at \mbox{$\omega = 90^\circ$}, or \mbox{$\Omega = 12^\circ\!\!.8$},
when \mbox{$i \!=\!  180^\circ \!-\!  B_\pi \!=\!  144^\circ\!\!.8$} from
Eq.\,(2).

\section{Interpreting the Variations in the Latitude of Perihelion}
Analysis in Sec.~4 of the variations in the angular orbital elements and
in the orientation of the line of apsides of the dwarf Kreutz sungrazers'
orbits has shown that the observed systematic increase in the latitude of
perihelion with the longitude of the ascending node in Figure 3 and in
Tables 2 and 3 is --- on the condition of an invariable longitude of
perihelion (also apparent from the figure and the two tables) --- a product
of some perturbations, or momentum increments, in the direction {\it
normal\/} to the comets' orbital planes.  This effect can in principle
be either continuous or discrete.  The magnitude of the effect is so profound
that, if discrete, it must consist of a large number of individual events,
because the separation velocities during a cometary splitting are known to
be at most only a few meters per second (e.g., Sekanina 1982, 2005), much
too low to fit \mbox{$\Delta B_\pi \simeq 25^\circ$}.

\subsection{Interpretation in Terms of a Sequence of Fragmentation Events}
The first scenario we consider is based on a hypothesis that the large
range of perihelion latitudes represents an accumulation of minor impulses
acquired by these dwarf comets during the many fragmentation events in the
course of one revolution about the Sun, from perihelion to next perihelion.

To scrutinize this hypothesis in some detail, we need to postulate a law
governing the sequence of such fragmentation events and to estimate an
average normal component of the impulse (differential velocity) a fragment
ought to acquire per event in order to match the total observed effect.
For this purpose we adopt the fragmentation sequence proposed by Sekanina
(2002), which was found to be in fair agreement with the known sequence of
secondary breakups of comet D/1993~F2 (Shoemaker-Levy~9).  The time $t_m$
of an $m$-th event is given by 
\begin{equation}
t_m = t_0 + {\displaystyle \frac{\Gamma^m - 1}{\Gamma - 1}} \Delta t,
\end{equation}
where $t_0$ is the time of the primary fragmentation event (close to
perihelion), which is tidal or tidally-supported or triggered in nature;
$\Delta t$ is an initial interval for secondary fragmentation; and
\mbox{$\Gamma > 1$} is a dimensionless constant that describes the rate
of fragmentation slowdown along the orbit.  If we reckon time from
perihelion, then \mbox{$t_0 \simeq 0$} and $\nu_0$, the total number
of fragmentation events or impulses imparted to a fragment, equals
\begin{equation}
\nu_0 = \frac{\log\left[ 1 + (\Gamma - 1) P_{\rm orb} / \Delta t \right]}{\log
 \Gamma},
\end{equation}
where $P_{\rm orb}$ is the orbital period.  This equation is a relation
between $\nu_0$, $\Gamma$, and $\Delta t$.  If, as was argued above, the
dwarf Kreutz sungrazers in the thickly populated branch in Figure 1 are
closely related to comet C/1843~D1 and if this comet, as postulated in
\mbox{Sekanina} \& Chodas (2007), is indeed the most massive known fragment
of the celebrated sungrazer X/1106~C1, then the dwarf Kreutz sungrazers
have orbital periods close to \mbox{$P_{\rm orb} = 900$}~years.

Next, it is necessary to adopt a certain breakup pattern.  Let us first
consider that a Kreutz fragment born from the primary breakup at the Sun
splits into two approximately equal pieces, each of which again breaks up
into two about equal parts, etc., until the fragments are eventually as
small {\vspace{-0.04cm}}as the faint dwarf sungrazers.  Assuming
10$^{18}$\,grams for the mass ${\cal M}_{\rm init}$ of the initial
fragment and 10$^6$\,grams for an average mass ${\cal M}_{\rm fin}$ of
the final products, the number of fragmentation events $\nu_0$ must
satisfy a condition \mbox{${\cal M}_{\rm fin} = 2^{-\nu_0} {\cal M}_{\rm
init}$}, so that in this case \mbox{$\nu_0 \simeq 40$}.  Equation (18) is
now a relation between $\Gamma$ and $\Delta t$.  For example, for $\Gamma$
between 1.1 and 1.2 the initial interval $\Delta t$ drops from 2~years
to 45~days.

It is now possible to apply Equation (13) to the proposed sequence of
fragmentation events, described by pairs of various values of $\Gamma$
and $\Delta t$.  Expressing $\Delta B_\pi$ in degrees, $p$ and $r$ in AU,
and $V_{\rm N}$ in m\,s$^{-1}$, the reciprocal gravitational constant
\mbox{$1/k = 0^\circ\!.001924$}.  The product $|r \sin u|$, the variable
in Eq.\,(13) for $\Delta B_\pi$, reaches a peak value
\begin{equation}
|r \sin u|_{\rm peak} = q (1 \!+\! e)^{\frac{1}{2}}(1 \!-\! e)^{-\frac{1}{2}}
 = a_0 (1 \!-\! e^2)^{\frac{1}{2}}
\end{equation}
at a distance $r_{\rm peak}$ that is equal to the semimajor axis $a_0$ of the
orbit.  Adopting \mbox{$q \simeq 0.0055$}~AU and \mbox{$e \simeq 0.99994$}, or
\mbox{$a_0 = 91.7$}~AU, we find \mbox{$|r \sin u|_{\rm peak} \simeq 1.00$}~AU
and for any single event \mbox{$|\Delta B_\pi| \ll 0^\circ\!.04\,V_{\rm N}$}.

Since the normal component of the separation velocity of fragments of the
split comets does not exceed a few meters per second (see the first paragraph
of Sec.~5) and, in addition, the sign of the expression for $\Delta B_\pi$
changes at aphelion (because of $\sin u$), so that \mbox{$\Sigma \Delta B_\pi
\ll \Sigma |\Delta B_\pi|$}, we conclude that this scenario cannot explain
the magnitude of the systematic rate of change in the latitude of perihelion
with the longitude of the ascending node in Tables~2--3 and Figure~3.  This
holds true even if we adopt a different breakup pattern with a much greater
total number of fragmentation events and/or a different law governing their
sequence.

This information is consistent with the perturbations due to a fragment's
separation velocity, listed in Table 8 of Sekanina (2002).  While a normal
velocity of 5~m\,s$^{-1}$ can trigger a change of up to nearly 27$^\circ$ in
the longitude of the nodal line and up to nearly 5$^\circ$ in the inclination,
it fails to shift the latitude of perihelion by even 0$^\circ\!$.1.

\subsection{Interpretation in Terms of Perturbations by\\a Major
 Nongravitational Force Near the Sun}
This category of hypotheses regards sizable deviations of the lines of
apsides of dwarf Kreutz sungrazers from the line of reference apsidal
orientation (Figure 3 and Tables 2 and 3) to be an effect of a continuous
force, which is nongravitational in nature and acts while these sungrazers
are under observation.  We first discuss briefly the Lorentz force, which
acts in the direction normal to the orbital plane.

The Lorentz force is known to appreciably affect the motions of
submicron-sized charged particles of dust.  Compelling evidence suggests,
however, that the dust tails of the dwarf Kreutz sungrazers,
consisting of such microscopic grains, are subjected to no detectable
effects of the Lorentz force (Sekanina 2000, Thompson 2009).  On the
strength of this argument, it is inconceivable that the nuclei of the
dwarf Kreutz sungrazers could be subjected to this force to an extent
of triggering a major effect of the kind examined.

An alternative scenario is based on an assumption that these are effects
due to a momentum transferred to the nucleus by sublimation of water ice
{\it and/or\/} more refractory species at small heliocentric distances,
at which the dwarf Kreutz sungrazers are observed.  Such small heliocentric
distances are strongly suggested by a peak inclination near \mbox{$\Omega
\simeq 340^\circ$} in Figure~1.  This value of $\Omega$ corresponds to
\mbox{$\omega \simeq 63^\circ$} in Figure~2.  Furthermore, the peak
inclination requires that \mbox{$\cot(\omega\!+\! u) = 0$} in Eq.\,(15),
so that \mbox{$u \simeq -153^\circ$}, equivalent to a time a little less
than 1~day before perihelion, when the dwarf sungrazers are indeed
observed.

In order for this scenario to work --- that is, to explain the effect normal
to the orbital plane --- we need to test whether the introduction into
the equations of motion of a term containing a normal component of the
momentum-transfer acceleration could offer orbital solutions in which the
angular elements are consistent with the reference apsidal orientation
presented in Table~1.

The standard orbit determination technique, which was developed
by Marsden et al.\ (1973) and whose versions are nowadays almost
universally employed worldwide to compute cometary orbits, sets up
the nongravi\-tational terms in the three cardinal directions of an RTN
right-handed coordinate system,\footnote{RTN = Radial/Transverse/Normal.}
identical to that in Sec.\ 4.  The introduced law, $g_{\rm ice}(r)$
(see Sec.~7), mimicks the dependence of an averaged water-ice sublimation
rate on heliocentric distance $r$.  The magnitude of the momentum-transfer
acceleration at 1~AU from the Sun is measured by the so-called Style~II
parameters, $A_1$ in the radial direction (away from the Sun), $A_2$ in
the transverse direction in the orbital plane, and $A_3$ in the direction
normal to the orbital plane {\vspace{-0.03cm}}(Marsden et al.\ 1973).
 Their values are tabulated in units of 10$^{-8}$\,AU~day$^{-2}$,
sometimes without an explicitly listed exponent.\footnote{A unit of
\mbox{10$^{-8}$~AU day$^{-2} = 2.004\!\times\!10^{-5}$\,cm s$^{-2}$}; at
1 AU from the Sun it equals $3.38 \times 10^{-5}$ the Sun's gravitational
acceleration.} For practically all comets whose motions have required the
incorporation of the nongravitational terms into the equations of motion,
successful orbital solutions have almost always been obtained by ignoring
the acceleration's normal component (\mbox{$A_3 \!=\! 0$}).  Marsden \&
Williams' (2008) catalogue of orbits does not even have a column for $A_3$.

To examine the proposed scenario, we first selected a small set of
dwarf Kreutz sungrazers from a catalogue of about 1600 of them (Marsden
\& Williams 2008; plus the objects from mid-May 2008 to mid-June
2010\,\footnote{These orbital elements have been published in numerous MPCs
starting with MPC\,63377 and ending with MPC\,72855.}).  Each selected
dwarf sungrazer was to satisfy these conditions:\ (i)~be a member
of the thickly populated branch in Figure 1; (ii)~the value of $\Omega$ to
come from a broad range of the longitudes of the ascending node, between
$\sim$300$^\circ$ and $\sim$40$^\circ$ (Table 2); (iii) as published, the
parabolic orbit be consistent with a least-squares solution, with no element
forced to any particular value; and (iv)~the orbit be based on at least
seven astrometric positions, measured from the images taken with the C2
coronagraph on board the {\it SOHO\/} satellite and/or with one of the COR2
coronagraphs on board the two {\it STEREO\/} spacecraft.  This last condition
stems from our focus on the final segment of the trajectory and from our need
to have astrometric observations of the best possible quality.  Because of
the pixel sizes, 11$^{\prime\prime}\!$.4 for the C2 coronagraph and
14$^{\prime\prime}\!$.7 for the COR2 coronagraphs, more accurate data were
expected from them than from the wide-field coronagraphs, C3
(56$^{\prime\prime}$/pixel) and HI1 (70$^{\prime\prime}$/pixel).  However,
the pixel-size advantage of C2 may sometimes be offset by scarcity or a very
uneven distribution of reference stars over the image field.

\begin{table*}[t]
\vspace{0.1cm}
\begin{center}
{\footnotesize {\bf Table 4}\\[0.1cm]
{\sc Comparison of Gravitational Solutions for Eight Test Dwarf Kreutz
Sungrazers\\with Their Nongravitational Solutions Containing $A_3$.}\\[0.1cm]
\begin{tabular}{c@{\hspace{0.3cm}}l@{\hspace{0.52cm}}l@{\hspace{0.18cm}}c@{\hspace{0.5cm}}c@{\hspace{0.4cm}}c@{\hspace{0.4cm}}c@{\hspace{0.3cm}}c@{\hspace{0.3cm}}c@{\hspace{0.3cm}}c@{\hspace{0.35cm}}c@{\hspace{0.3cm}}c@{\hspace{0.3cm}}c}
\hline\hline\\[-0.2cm]
 & & & \multicolumn{5}{@{\hspace{-0.2cm}}c}{Orbital elements (eq.\ J2000)}
 & Param- & \multicolumn{2}{@{\hspace{-0.15cm}}c}{Line of apsides} & Apsidal
 & $N_{\rm obs}$; \\[-0.04cm]
 & & Orbital  & \multicolumn{5}{@{\hspace{-0.1cm}}c}{\rule[0.6ex]{6.8cm}{0.4pt}}
 & eter\,$A_3$ & \multicolumn{2}{@{\hspace{-0.15cm}}c}{\rule[0.6ex]{2.15cm}{0.4pt}}
 & line's & \raisebox{-0.07cm}{RMS} \\[-0.04cm]
No. & Object & solution\rlap{$^{\rm a}$} & $t_\pi$\,(ET) & $\omega$ & $\Omega$
    & $i$ & $q$\,({\Rssun}) & (units$^{\rm b}$) & $L_\pi$ & $B_\pi$ & offset
    & residual \\[0.08cm]
\hline \\[-0.18cm]
1 & C/2007 X13 & Cat. & 2007/12/14.42 & 27$^\circ\!\!$.73 & 305$^\circ\!\!$.98
    & 138$^\circ\!\!$.31 & 1.63 & \ldots\ldots.\,. & 284$^\circ\!\!$.55
    & +18$^\circ\!\!$.03 & 17$^\circ\!\!$.24 & $\;\:$8 \\
  & & Grav. & 2007/12/14.42 & 27.76 & 306.00 & 138.31 & 1.62 & \ldots\ldots.\,.
    & 284.54 & +18.05 & 17.22 & $\pm$12$^{\prime\prime}\!$.7 \\
  & & ($A_3$) & 2007/12/14.53 & 87.33 & $\;\:\;\:$9.69 & 144.87 & 1.03
    & \llap{$-$}25.1$\!\!\!\:$ & 282.95 & +35.08 & $\;\:$0.17
    & $\pm$13$^{\prime\prime}\!$.0 \\[0.1cm]
2 & C/2007 X3 & Cat. & 2007/12/05.16 & 39.53 & 315.29 & 142.81 & 1.37
    & \ldots\ldots.\,. & 281.97 & +22.63 & 12.59 & $\;\:$7 \\
  & & Grav. & 2007/12/05.16 & 39.55 & 315.30 & 142.81 & 1.38 & \ldots\ldots.\,.
    & 281.96 & +22.64 & 12.58 & $\;\:\pm$5$^{\prime\prime}\!$.6 \\
  & & ($A_3$) & 2007/12/05.25 & 82.08 & $\;\;\:\:$4.55 & 145.18 & 1.33
    & \llap{$-$}14.8$\!\!\!\:$ & 284.17 & +34.44 & $\;\:$1.36
    & $\;\:\pm$6$^{\prime\prime}\!$.6 \\[0.1cm]
3 & C/2001 Y4 & Cat. & 2001/12/18.62 & 42.87 & 324.05 & 144.61 & 1.50
    & \ldots\ldots.\,. & 286.93 & +23.20 & 12.53 & 10 \\
  & & Grav. & 2001/12/18.62 & 42.86 & 324.03 & 144.60 & 1.51 & \ldots\ldots.\,.
    & 286.93 & +23.20 & 12.53 & $\pm$20$^{\prime\prime}\!$.0 \\
  & & ($A_3$) & 2005/12/18.68 & 86.79 & $\;\:\;\:$9.22 & 144.68 & 1.02
    & $-$4.1\rlap{8} & 283.15 & +35.26 & $\;\:$0.29
    & $\pm$20$^{\prime\prime}\!$.8 \\[0.1cm]
%
% 4 & C/2009 X15 & Cat. & 2009/12/14.07 & 61.01 & 338.27 & 145.40 & 1.10
%   & \ldots\ldots & 282.22 & +29.78 & $\;\:$5.44 & 11 \\
% & & Grav. & 2009/12/14.07 & 61.00 & 338.25 & 145.40 & 1.09 & \ldots\ldots
%   & 282.21 & +29.78 & $\;\:$5.44 & $\;\:\pm$9$^{\prime\prime}\!$.5 \\
% & & ($A_3$) & 2009/12/14.10 & 82.46 & $\;\:\;\:$1.87 & 144.53 & 0.97
%   & $-$3.3\rlap{5} & 281.10 & +35.03 & $\;\:$1.40
%   & $\;\:\pm$9$^{\prime\prime}\!$.6 \\[0.1cm]
%
4 & C/2008 M4 & Cat. & 2008/06/25.69 & 58.09 & 338.00 & 144.63 & 1.29
    & \ldots\ldots.\,. & 285.37 & +29.43 & $\;\:$6.16 & 10 \\
  & & Grav. & 2008/06/25.69 & 58.01 & 337.92 & 144.62 & 1.29 & \ldots\ldots.\,.
    & 285.37 & +29.41 & $\;\:$6.19 & $\;\:\pm$8$^{\prime\prime}\!$.5 \\
  & & ($A_3$) & 2008/06/25.71 & 79.00 & 358.94 & 144.16 & 1.00
    & $-$3.0\rlap{4} & 282.43 & +35.08 & $\;\:$0.33
    & $\;\:\pm$8$^{\prime\prime}\!$.8 \\[0.1cm]
5 & C/2009 L5 & Cat. & 2009/06/05.32 & 75.97 & 355.06 & 144.31 & 1.01
    & \ldots\ldots.\,. & 282.16 & +34.47 & $\;\:$0.90 & 12 \\
  & & Grav. & 2010/06/05.32 & 75.77 & 354.81 & 144.34 & 1.01 & \ldots\ldots.\,.
    & 282.14 & +34.41 & $\;\:$0.96 & $\;\:\pm$6$^{\prime\prime}\!$.8 \\
  & & ($A_3$) & 2010/06/05.33 & 78.85 & 358.67 & 143.94 & 1.02 & $-$0.4\rlap{9}
    & 282.37 & +35.28 & $\;\:$0.36 & $\;\:\pm$6$^{\prime\prime}\!$.9 \\[0.1cm]
6 & C/2006 J9 & Cat. & 2006/05/10.98 & 86.59 & $\;\:\;\:$9.93 & 143.37 & 1.18
    & \ldots\ldots.\,. & 284.18 & +36.55 & $\;\:$1.75 & 13 \\
  & & Grav. & 2006/05/10.98 & 86.58 & $\;\:\;\:$9.92 & 143.38 & 1.18
    & \ldots\ldots.\,. & 284.18 & +36.54 & $\;\:$1.75
    & $\;\:\pm$3$^{\prime\prime}\!$.9 \\
  & & ($A_3$) & 2006/05/10.97 & 85.59 & $\;\:\;\:$6.94 & 144.00 & 1.06
    & +0.8\rlap{3} & 282.39 & +35.88 & $\;\:$0.76
    & $\;\:\pm$3$^{\prime\prime}\!$.9 \\[0.1cm]
7 & C/2008 M5 & Cat. & 2008/06/26.40 & 99.81 & $\;\:$22.08 & 139.96 & 0.90
    & \ldots\ldots.\,. & 279.35 & +39.34 & $\;\:$4.97 & 10 \\
  & & Grav. & 2008/06/26.40 & 99.69 & $\;\:$21.93 & 140.02 & 0.91
    & \ldots\ldots.\,. & 279.37 & +39.30 & $\;\:$4.92
    & $\;\:\pm$9$^{\prime\prime}\!$.6 \\
  & & ($A_3$) & 2008/06/26.37 & 82.38 & $\;\:\;\:$3.35 & 144.59 & 1.08
    & +3.8\rlap{4} & 282.67 & +35.06 & $\;\:$0.18
    & $\;\:\pm$9$^{\prime\prime}\!$.2 \\[0.1cm]
8 & C/2008 K8 & Cat. & 2008/05/28.65 & 99.02 & $\;\:$32.42 & 135.31 & 1.72
    & \ldots\ldots.\,. & 289.83 & +43.99 & 10.32 & 14 \\
  & & Grav. & 2008/05/28.65 & 98.96 & $\;\:$32.19 & 135.44 & 1.71
    & \ldots\ldots.\,. & 289.71 & +43.88 & 10.18
    & $\pm$10$^{\prime\prime}\!$.2 \\
  & & ($A_3$) & 2008/05/28.53 & 84.47 & $\;\:\;\:$5.59 & 144.45 & 1.01
    & +3.2\rlap{5} & 282.38 & +35.36 & $\;\:$0.38
    & $\pm$10$^{\prime\prime}\!$.3 \\[0.05cm]
\hline \\[-0.2cm]
\multicolumn{13}{l}{\parbox{16.7cm}{$^{\rm a}$\,{\scriptsize
 Cat.\,=gravitational orbit from catalogue by{\vspace*{-0.06cm}} Marsden
 \& Williams (2008); Grav.\,=\mbox{gravitational orbit computed by
 us};~($A_3$)\,=\,nongravitational orbit with $A_3$ forced to fit the
 line of reference apsidal orientation as closely as possible.}}}\\[-0.05cm]
\multicolumn{13}{l}{\parbox{16.7cm}{$^{\rm b}$\,{\scriptsize Units of
 10$^{-5}$AU\,day$^{-2}\!$, a thousand times greater than units used in
 standard orbital computations; normalized to 1 AU from the Sun.}}} \\
\end{tabular}}
\end{center}
\end{table*}

All orbital computations were carried out by the second author, who employed
a code {\it EXORB7\/} developed by A.\ Vitagliano.  The code includes the
perturbations by the eight major planets, Pluto, and the three most massive
asteroids.  It employs the standard DE406 library and allows one to use a
forced value for  any orbital element or nongravitational parameter, an
option that was copiously exploited.  For each of the selected dwarf Kreutz
sungrazers, two sets of orbital elements were derived.  The first set was
a parabolic gravitational solution, whereas the second set was a restricted
nongravitational solution, which employed a parabolic approximation and
Marsden et al.'s (1973) standard formalism, in which we assumed that
\mbox{$A_1 \!=\! A_2 \!=\! 0$} and which we used to search for the magnitude
of the acceleration component normal to the orbit plane, \mbox{$A_3 \neq 0$}.
Because of fairly large uncertainties in the astrometric observations and
very short orbital arcs observed, we decided at this point not to incorporate
$A_3$ as a variable directly into the least-squares differential optimization
procedure, but, instead, to proceed by iteration.

Based on our estimates of several meters for the nuclear sizes of faint
dwarf Kreutz sungrazers near the end of their lifetime (by
extrapolating the derived diameters of brighter ones; e.g., Sekanina
2003) and on the highest known values of \mbox{$A_1\!\sim\!20 \!-\!30\!
\times\!10^{-8}$}\,AU\,day$^{-2}$ (Marsden\,\&\,Williams 2008) among
ordinary long-period comets (in particular, objects such as C/1993~A1 or
C/1998~P1) with presumably kilometer- or subkilometer-sized nuclei, we
began each computer run with an initial value of \mbox{$A_3 \approx
10^{-5}$}\,AU\,day$^{-2}$.  This was a fairly conservative estimate since the
comparison comets, of perihelion distances comparable to or exceeding 1~AU,
imply momentum-transfer effects due almost exclusively to water ice driven
sublimation, while at the heliocentric distances at which the dwarf
Kreutz sungrazers are observed --- around 0.05~AU or $\sim$11~{\Rsun} ---
numerous species less volatile than water ice also sublimate profusely.
Together with progressive fragmentation this should increase the magnitude
of the momentum-transfer acceleration.  Thus, applying a least-squares
optimization procedure to derive the orbital elements, without removing
any of the astrometric positions available, we searched for a value of
$A_3$ that, in the context of the common origin of the Kreutz sungrazers
(Sec.~1), provided a minimum offset from the reference apsidal orientation.

The results for eight test dwarf Kreutz sungrazers are summarized in
Table~4, in which we compare three sets of orbital solutions.  The
first set, in the row {\it Cat.\/}, is the orbit as computed by Marsden;
it is copied either from Marsden \& Williams' (2008) catalogue (entries
1--3 and 6), or from MPC\,63599--63601 (entries 4 and 7--8) or MPC\,66704
(entry 5).

The second set, in the row {\it Grav.\/}, represents our own parabolic
gravitational approximation.  We made this run in order to confirm that
Marsden's results, the details of which have never been published, are
closely reproduced.  Comparison with the catalogued orbital elements
tolerated formal differences of up to a few tenths of a degree in the
angular elements and up to 0.01\,{\Rsun} in the perihelion distance.  As
expected, the perihelion times always agreed to better than 0.01~day.
However, a number of objects, especially from the earlier times, considered
initially as suitable test cases, had to be rejected, because the
catalogued perihelion distances were forced by Marsden, usually upwards
but sometimes downwards, to make them exceed 1\,{\Rsun} but stay smaller
than $\sim$2\,{\Rsun}.  This manipulation resulted in changes in the other
elements as well, including the longitude of the ascending node, and
consequently interfered with our effort to have the test sungrazers
distributed more or less uniformly between \mbox{$\Omega \simeq 300^\circ$}
and \mbox{$\Omega \simeq 40^\circ$}.

The third set, in the row ($A_3$), is our nongravitational solution,
referred to above.  The deviation of an iterated value of $A_3$ from the
value we were searching for was measured by an offset of the iterated apsidal
orientation, given by the perihelion coordinates $L_\pi$ and $B_\pi$,
\begin{eqnarray}
\tan (L_\pi - \Omega) & = & \tan \omega \, \cos i, \nonumber \\[-0.24cm]
 & & \\[-0.28cm]
\sin B_\pi & = & \sin \omega \, \sin i, \nonumber
\end{eqnarray}
from the reference orientation.  The offset $\epsilon$, whose minimum we
were aiming at, was computed from
\begin{equation}
\cos \epsilon = \sin B_\pi \sin \langle B_\pi \rangle + \cos B_\pi
 \cos \langle B_\pi \rangle \cos (L_\pi \!-\! \langle L_\pi \rangle).\\[0.1cm]
\end{equation}
The iteration proceeded by trial and error until for three chosen values of
$A_3$, relatively close to one another and preferably equidistant or nearly
equidistant, and such that \mbox{$(A_3)_1 < (A_3)_2 < (A_3)_3$}, the
respective offsets, $\epsilon_1$, $\epsilon_2$, $\epsilon_3$, from the
reference apsidal line satisfied a condition \mbox{$\epsilon_2 < \min
(\epsilon_1, \epsilon_3)$}.  A parabola{\vspace{-0.02cm}} was then fitted
through the three $\epsilon^2(A_3)$ points, with the{\vspace{-0.03cm}}
square root of the minimum $\epsilon_{\rm min}^2$ listed in column 12 and
the resulting $A_3$ value in column 9 of the ($A_3$) row in~\mbox{Table}~4.
The orbital elements, listed in columns 4--8 of the same row, were then
recomputed with this value of $A_3$.

Together with Figures 5 and 6, in which we plot, respectively, the tested
sungrazers' inclination and argument of perihelion against their longitude of
the ascending node, Table~4 allows us to make a number of conclusions.
Most importantly, for every single tested object the {\it introduction of
a~nongravitational solution leads to a dramatic drop in the offset from the
reference apsidal line\/} and the {\it derived angular orbital elements are
quite unlike those from the gravitational solution\/}, the magnitude of the
differences correlating with the $A_3$ parameter.  The longitude of the
ascending node and the argument of perihelion changed by as much as
$\sim$60$^\circ$~(sic!), the inclination by up to 9$^\circ$, the perihelion
distance by as much as 0.7\,{\Rsun}, and even the perihelion time by up
to 0.12~day.  The range of orbital differences among the objects was
reduced considerably by the nongravitational solutions, from 86$^\circ$ to
11$^\circ$ in $\Omega$, from 72$^\circ$ to 8$^\circ\!$.5 in $\omega$,
from 9$^\circ$ to 1$^\circ\!$.2 in $i$, and from 0.8~{\Rsun} to 0.33~{\Rsun}
in $q$.

The minimum offset from the reference apsidal line offered by the
nongravitational solutions for the eight sungrazers varies from less than
0$^\circ\!$.2 to almost 1$^\circ\!$.4.  Even in the least favorable case
is the offset reduced by more than a factor of two compared to the
gravitational solution, while more typically the reduction factor is
between 10 and 100.  And for the seven cases with offsets smaller than
1$^\circ$, the perihelion distance is confined to a narrow range from
1.00~{\Rsun} to 1.08~{\Rsun}.

\begin{figure}[t]
\vspace*{0.08cm}
\hspace*{-0.62cm}
\centerline{
\scalebox{0.492}{
\includegraphics{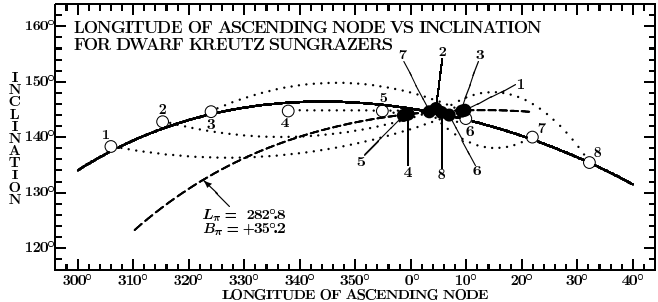}}} % from f5_SOHO-rev.tex
\vspace*{-8.45cm}
\caption{Plot of the longitude of the ascending node against the orbit
inclination for eight test dwarf Kreutz sungrazers.  The open
circles refer to the gravitational solutions (rows Grav.\ in Table 4),
the solid circles to the nongravitational solutions in rows ($A_3$).  The
number at each circle identifies the sungrazer via column 1 in Table 4.
The thick dashed curve is a predicted locus of comets whose orbits fit
the reference apsidal line, while the thinner solid curve is its rotated
version (curve A in Figure 4).  The dotted curves only serve to connect
the two types of solution for each object (to show the enormity of the
corrections especially in the longitude of the ascending node); their
shape is arbitrary.{\vspace{0.25cm}}}
\end{figure}

Our a priori estimate of 10$^{-5}$\,AU day$^{-2}$ for the parameter $A_3$
compares rather favorably with the results.  It is within a factor of $\sim$4
for six of the eight entries in Table~4 and is closer to the lower end of the
range.  The $A_3$ values statistically correlate both with the offsets from
the line of apsides and with the longitudes of the ascending node derived from
the gravitational solutions.  However, as confirmed by our additional tests,
there is no functional correlation.  The crossover from negative to positive
values of $A_3$ is near the position of comet C/1843~D1.

Of much interest are the magnitudes of $|A_3|$ at the upper end of their range.
The value for comet C/2007~X13 (normalized to a distance of 1~AU from the Sun)
is equivalent to 0.503~cm~s$^{-2}$, alarmingly close to 0.593~cm~s$^{-2}$, the
Sun's gravitational acceleration.  At a heliocentric distance of about
0.046~AU, at which the comet's image was last astrometrically measured, the
Sun's gravitational acceleration is 280~cm~s$^{-2}$, or 0.140 AU day$^{-2}$,
while the applied nongravitational law predicts for this distance a water-ice
sublimation rate 768~times greater than at 1~AU from the Sun,\footnote{The
nongravitational law predicts that at small heliocentric distances the
water-ice sublimation rate (and the sungrazer's momentum-transfer
acceleration) varies at a rate slightly steeper than the Sun's gravitational
acceleration (Sec.~7).} so that the nongravitational acceleration at 0.046~AU
was 387~cm~s$^{-2}$ or 0.193~AU~day$^{-2}$.  The results in Table~4 thus
imply that at the end of its visible trajectory, {\it C/2007~X13 was
subjected to a normal component of the nongravitational acceleration that
exceeded the Sun's gravitational acceleration by nearly 40~percent!\/}
For the other objects in Table~4 the numbers are less extreme but still
remarkable.

We should also comment on the relative magnitude of the RMS residuals from
the orbital solutions in the last column of Table 4.  Not in a single case
is the formal fit from the nongravitational solution markedly worse than from
the gravitational solution.  Thus, the introduction of $A_3$ was tolerable
from the data-analysis standpoint; it was of course vital from the standpoint
of our dynamical arguments.  The equivalence of the gravitational and
nongravitational solutions in terms of the RMS residuals is not surprising;
it is simply a sign of fairly low accuracy of the astrometric observations.
It is the offset from the apsidal line and not the formal quality of fit
that is the {\it driver\/} in our undertaking this task and the {\it prime
criterion\/} in measuring the significance of the results.

\begin{figure}[t]
\vspace*{0.08cm}
\hspace*{-0.61cm}
\centerline{
\scalebox{0.49}{
\includegraphics{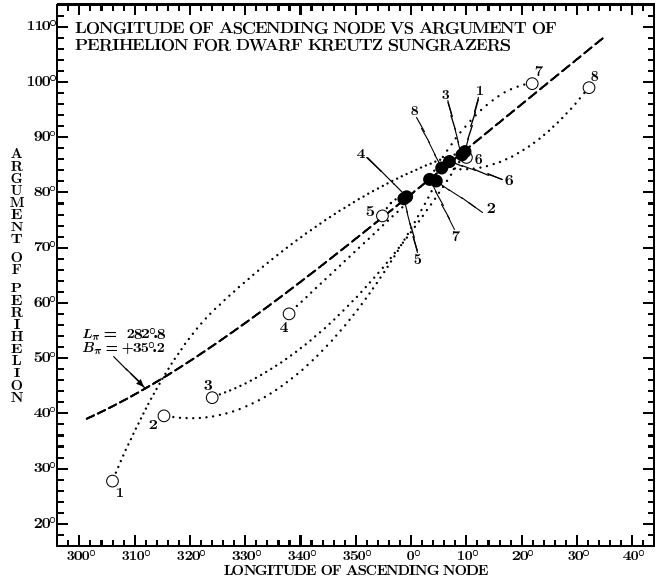}}} % from f6_SOHO-rev.tex
\vspace*{-4.3cm}
\caption{Plot of the longitude of the ascending node against the argument
of perihelion for eight test dwarf Kreutz sungrazers. For the
description see the caption to Figure 5.{\vspace*{0.35cm}}}
\end{figure}

Graphically, the reduction of the ranges of the three angular elements
$\Omega$, $i$, and $\omega$ brought about by the introduction of the
nongravitational term with the parameter $A_3$, is dramatically revealed
by Figures~5 and 6.  From Figure~5 it is obvious that all test sungrazers
whose gravitational solutions were distributed along the thickly populated
branch associated with C/1843~D1 are in their nongravitational solutions
distributed tightly along the line of apsides instead. All shifts in the
test objects' positions in the plot of $\Omega$ against $i$, however large,
were accomplished by ``sliding'' along, not across, the curve.

Our last comment on the results in Table 4 and Figures~5 and 6 is to
point out the unanswered questions and employed approximations.  Although
the nongravitational solutions for the eight test sungrazers appear to
be satisfactory and are responsive to the argument based on the observed
trend in $B_\pi$ from gravitational solutions, the effects driven by
the process of erosion should, in general, include a radial component
(parameter $A_1$) and a transverse component (parameter $A_2$) as well.
While the two components in the orbital plane do not affect the
longitude of the ascending node or the inclination directly, the
introduction into the equations of motion of their contributions could
affect the value of $A_3$ as well.  We return to this issue, which
requires a fuller incorporation of erosion-driven effects, in Sec.~8.

\begin{figure}[t]
\vspace{0.22cm}
\hspace{-1.23cm} % -1.18
\centerline{
\scalebox{0.54}{ % 0.51
\includegraphics{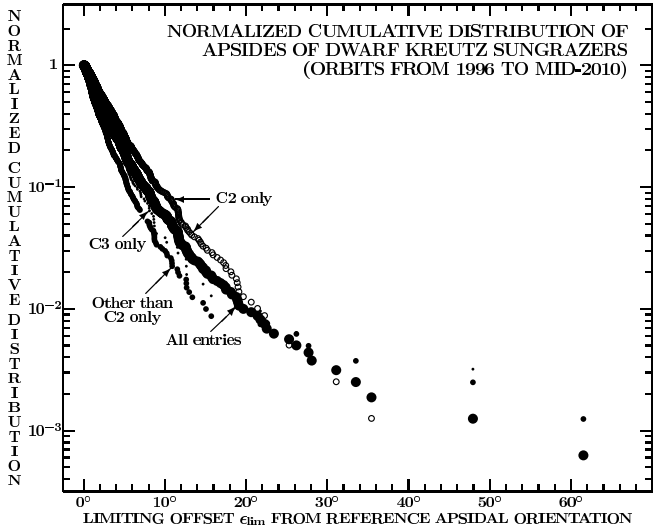}}} % f7_SOHO-rev.tex
\vspace*{-6.35cm} % -5.9
\caption{Normalized cumulative distribution of dwarf Kreutz sungrazers
with offsets from the reference apsidal orientation equal to or exceeding
a limiting offset $\epsilon_{\rm lim}$.  Four sets of the dwarf sungrazers
are plotted:\ (a)~a complete set of about 1600 entries (large bullets,
marked:\ {\it All entries\/}); (b)~sungrazers with all astrometric positions
from images taken only with the C2 coronagraph on board {\it SOHO\/} (open
circles, marked:\ {\it C2 only\/}); (c)~sungrazers with positions from images
taken with any {\it SOHO\/} or {\it STEREO\/} coronagraph other than C2 only
(small bullets, marked:\ {\it Other than C2 only\/}); and (d)~sungrazers with
all positions from images taken only with the C3 coronagraph on board {\it
SOHO\/} (dots, marked:\ {\it C3 only\/}).{\vspace*{0.35cm}}}
\end{figure}

\section{Apparent Scatter in Apsidal Orientation\\from Gravitational Orbits
 of\\Dwarf Kreutz Sungrazers}

An important question to answer is whether the enormous scatter in the angular
orbital elements of the dwarf Kreutz sungrazers derived from
the gravitational solutions is only a dynamical effect, as discussed in the
previous section or whether there also is a significant contribution from
the errors of measurement and/or reduction of the astrometric observations.
To address this question, the plot in Figure 7 shows, as a function of a
limiting offset $\epsilon_{\rm lim}$, a cumulative distribution of these
objects whose angular offsets $\epsilon$ from the reference apsidal
orientation exceed, or are equal to, $\epsilon_{\rm lim}$.

Figure 7 and Table 5, which summarizes the results from the cumulative
distribution, show that the offsets from the reference apsidal
orientation are the largest for the set of sungrazers observed with the C2
coronagraph only.  This suggests that the gravitational orbits of the dwarf
sungrazers are the most inaccurate when they are based only on astrometric
observations from C2 images.  This effect could be instrumental, because
the orbital arcs in C2 are shorter and the number of reference stars is
sometimes insufficient or their field distribution strongly nonuniform,
both of which may lead to inferior orbital solutions.  On the other hand,
due to the much smaller pixel size of the C2 coronagraph, the astrometry
should be better. Perplexingly, if the greater offsets in Figure~7 were an
instrumental effect, one would expect that the {\it C3 only\/} distribution
curve should lie below the {\it Other than C2 only\/} distribution, because
the latter includes largely the sungrazers whose orbits were based on images
from both C3 and C2.  This expectation is however contradicted in the figure,
as the {\it C3 only\/} distribution shows that the orbits based on C3 images
yield offsets that are larger than those from the orbits based on C3 {\it
plus\/} C2, so that C2 images improve the orbits.

We suggest that the solution to this puzzle lies in the same issue that
led us to examine a set of representative examples of sungrazers observed
{\it only\/} in C2, that is, at the very end of their trajectory.  At this
late evolutionary phase, the size of the nucleus of a dwarf sungrazer is
near zero because of escalating erosion, and the deviation from a
gravitational motion is at a maximum.  Gravitational solutions are
inappropriate under these circumstances and their failure shows up as
an enormous effect in the apsidal orientation.

%
% The full incorporation of the erosion-driven effects in the motions of
% the SOHO/STEREO Kreutz sungrazers requires a more comprehensive orbital
% solution.  This issue is addressed next.
%
\begin{table}
\noindent
\vspace{-0.2cm}
\begin{center}
{\footnotesize {\bf Table 5}\\[0.1cm]
{\sc Cumulative Distribution of Apsidal Orientation Offsets of\\Dwarf Kreutz
 Sungrazers from Reference Apsidal\\Orientation
 (Eq.\ 2000).}\\[0.1cm]
\begin{tabular}{c@{\hspace{0.3cm}}c@{\hspace{0.85cm}}c@{\hspace{0.7cm}}c@{\hspace{0.7cm}}c}
\hline\hline\\[-0.25cm]
Offset from & \multicolumn{4}{@{\hspace{0.03cm}}c}{Fraction\,(\%)\,of
 dwarf sungrazers with}\\[-0.03cm]
reference apsidal  & \multicolumn{4}{@{\hspace{0.03cm}}c}{apsidal
 orientation offsets $\epsilon \!\ge\! \epsilon_{\rm lim}$}\\[-0.04cm]
orientation$^{\rm a}$,
 & \multicolumn{4}{@{\hspace{0.03cm}}c}{\rule[0.6ex]{5.73cm}{0.4pt}}\\[-0.03cm]
$\epsilon_{\rm lim}$ & All\rlap{$^{\rm b}$} & C2 only\rlap{$^{\rm c}$}
 & Other\rlap{$^{\rm d}$} & C3 only$^{\rm e}$ \\[0.12cm]
\hline\\[-0.15cm]
0$^\circ\!\!$.5 & 86 & 94 & 78 & 86 \\
1.0             & 75 & 83 & 58 & 71 \\
2.0             & 49 & 60 & 37 & 50 \\
4.0             & 26 & 35 & 17 & 24 \\
8.0             & $\;\:$9 & 27 & $\;\:$5 & $\;\:$7 \\
\llap{1}2.0     & $\;\:$3\rlap{.4} & $\;\:$5\rlap{.2} & $\;\:$1\rlap{.8}
                & $\;\:$2\rlap{.8} \\[0.04cm]
\hline\\[-0.23cm]
Totals$\;$      & 1598 & 794 & 804 & 313 \\[0.05cm]
\hline\\[-0.2cm]
\multicolumn{5}{l}{\parbox{8.3cm}{$^{\rm a}$\,{\scriptsize Standard apsidal
 direction{\vspace*{-0.03cm}} of Kreutz sungrazer system defined by:\
 $\langle L_\pi \rangle \!=\!  282^\circ\!.8$ and $\langle B_\pi \rangle \!=\!
 +35^\circ\!.2$.}}} \\[0.15cm]
\multicolumn{5}{l}{\parbox{8.3cm}{$^{\rm b}$\,{\scriptsize Set of
 dwarf Kreutz{\vspace*{-0.06cm}} sungrazers detected, measured, and
 reduced from images taken with any onboard coronagraph.}}} \\[0.22cm]
\multicolumn{5}{l}{\parbox{8.3cm}{$^{\rm c}$\,{\scriptsize Set of dwarf
 Kreutz{\vspace*{-0.06cm}} sungrazers detected, measured, and reduced only
 from images taken with C2 coronagraph on board {\it SOHO\/}.}}}\\[0.16cm]
\multicolumn{5}{l}{\parbox{8.3cm}{$^{\rm d}$\,{\scriptsize Set of
 dwarf{\vspace*{-0.06cm}}~Kreutz~\mbox{sungrazers}~\mbox{detected},~\mbox{measured},~and
 reduced from images taken with{\vspace*{-0.06cm}}
 coronagraphs~other~than~only~C2.}}} \\[0.25cm]
\multicolumn{5}{l}{\parbox{8.3cm}{$^{\rm e}$\,{\scriptsize Set of dwarf
 Kreutz{\vspace*{-0.06cm}} sungrazers detected, measured, and reduced only
 from images taken with C3 coronagraph{\vspace*{-0.05cm}} on board {\it
 SOHO\/}; this is a subset of the set `Other'.{\vspace{0.1cm}}}}}
\end{tabular}}
\end{center}
\end{table}

\section{Generalizing the Momentum-Transfer Law}

Our search for nongravitational orbital solutions in Sec.~5.2 used the Style
II formalism of Marsden et al.\ (1973); the standard nongravitational law,
incorporated into this formalism and approximating the sublimation rate of
water ice as a function of heliocentric distance, was employed by us
deliberately, because this convention allowed us to compare the magnitudes
of the nongravitational parameters that we determined for the dwarf Kreutz
sungrazers with the magnitudes for ordinary comets at much larger
heliocentric distances.  This comparison led to our discovery that the
ougassing-driven accelerations for the dwarf sungrazers were orders of
magnitude higher than those for ordinary comets and in the extreme cases
comparable in magnitude to the Sun's gravitational acceleration.

The standard momentum-transfer law, $g_{\rm ice}(r)$, has in Marsden et al.'s
(1973) formalism been expressed by an empirical formula,
\begin{equation}
g_{\rm ice}(r) = a \left( \:\!\!\frac{r}{r_0} \! \right)^{\!\!-m} \!\!\left[
 1 \!+\! \left( \:\!\!\frac{r}{r_0} \!\right)^{\!\!n} \right] ^{\!-k} \!\!,
\end{equation}
where $m = 2.15$, $n = 5.093$, $k = 4.6142$, the scaling distance \mbox{$r_0
= 2.808$}~AU, and the normalization constant \mbox{$a = 0.1113$} such that
\mbox{$g_{\rm ice}(1\,{\rm AU}) \!=\! 1$}.  These constants apply to a
so-called isothermal model of water-ice sublimation, which averages the Sun's
incident radiation over the surface of a spherical nucleus by assuming that
the temperature of the water-ice covered surface does not vary from site to
site and depends only on the heliocentric distance.  Although the temperature
varies over the surface greatly, for the orbit-determination purposes the
formula (22) has over the many decades provided excellent service and still
is employed nowadays.

Encouraged by the results summarized in Table 4, we continued our
experimentation with the nongravitational terms in the equations of motion by
testing nongravitational solutions for the dwarf Kreutz sungrazers observed
with the C2 coronagraph by directly incorporating the parameter $A_3$, next
to the orbital elements, as a variable into the least-squares optimization
procedure.  Employing Marsden et al.'s (1973) formalism, a few computer runs
confirmed our pessimism (Sec.~5.2) that this effort is doomed to failure, given
the uncertainties of the {\it SOHO\/} astrometry and very short orbital arcs of
the objects observed only with the C2 coronagraph.  We then tested this same
approach on extended orbital arcs available for dwarf Kreutz sungrazers
observed with both the C2 and C3 coronagraphs.  For a dwarf sungrazer C/2003~Q7,
for example,{\vspace{-0.04cm}} we found \mbox{$A_3=(+0.211 \pm 0.090) \times
10^{-5}$ AU day$^{-2}$} --- a marginal detection with a 1$\sigma$ relative
error of more than 40~percent --- by fitting the last 18 of the 45 astrometric
positions available.  On the other hand, a solution based on all 45 positions
led to a completely indeterminate result, \mbox{$A_3 = (+0.021 \pm 0.039)
\times 10^{-5}$ AU day$^{-2}$}.

We concluded that to solve for $A_3$ as one of the parametric variables
directly from the least-squares equations is not the way to proceed.
Nevertheless, this experimentation was helpful in that it suggested that
(i)~the value of $A_3$ had a tendency to increase by perhaps as much as one
order of magnitude as orbital solutions were derived from astrometric
observations closer to the point of the sungrazer's disappearance;
(ii)~the standard nongravitational law $g_{\rm ice}(r)$ is not quite
appropriate for the dwarf Kreutz sungrazers; and (iii)~laws that imply
a steeper variation with heliocentric distance are more compatible with
the astrometric observations.

Argument (iii) is strongly supported by the fact that numerous species
considerably less volatile than water ice, including atomic sodium (e.g.,
Knight et al.\ 2010), are known to sublimate profusely at heliocentric
distances smaller than $\sim$0.1~AU, where dwarf sungrazers are typically
observed.  It is therefore highly doubtful that the sublimation of water
ice typically dominates the erosion process in close proximity to the Sun.
Accordingly, the issue of an appropriate momentum-transfer law for the
dwarf sungrazers needs comprehensive examination.  If the momentum
transferred from the outgassing of other, more refractory species should
be more important than water ice, the scaling distance should be much
smaller than in Eq.\,(22), \mbox{$r_0 \ll 2.8$ AU}.

To accommodate a greater number of options and to make the momentum-transfer
law more flexible and realistic for applications to dwarf Kreutz sungrazers,
it is desirable to fundamentally generalize the momentum-transfer law.  In
this section we introduce three such laws.  One of them, $g_{\rm Na}(r)$,
is based on the sublimation of sodium (derived from the dependence of the
saturated vapor pressure on temperature; e.g., Hicks 1963), which is known to
outgas profusely from dwarf Kreutz sungrazing comets (e.g., Biesecker et al.\
2002; Sekanina 2003; Knight et al.\ 2010).  The resulting sublimation rate as
a function of heliocentric distance for the isothermal model can closely be
approximated by the formula (22) with the parameters:\ \mbox{$m = 2.089$},
\mbox{$n = 3.603$}, \mbox{$k = 4.896$}, \mbox{$r_0 = 0.3458$ AU}, and
\mbox{$a = 10^{9.145}$} (see also Sekanina \& Kracht 2014).  Although
the scaling distance $r_0$ is much smaller than for water ice, in a range
of heliocentric distances where the dwarf Kreutz sungrazers are typically
observed, sodium and water ice sublimate at rates that have a similar
dependence on $r$.  We would therefore expect that the two laws should
lead to similar results, a circumstance that is useful in checking the
validity of computations.

The second employed momentum-transfer law refers to one of very highly
refractory materials --- forsterite, the magnesium-rich end-member of
the olivine solid solution series (Mg$_2$SiO$_4$), a common silicate
in comets, including Kreutz sungrazers (e.g., Sekanina 2000, Kimura et
al.\ 2002, Ciaravella et al.\ 2010, Sekanina \& Chodas 2012).  From the
data of Hashimoto's (1990) laboratory experiments, we computed the
sublimation rate of forsterite as a function of heliocentric distance
in the isothermal case and found that it, too, can closely be
approximated by the expression (22) with the parameters:\ \mbox{$m =
2.634$}, \mbox{$n = 5.155$}, \mbox{$k = 3.320$}, \mbox{$r_0 =
0.014861\;{\rm AU} = 3.1926$}\,{\Rsun}, and \mbox{$a = 10^{36.10}$}
(see also Sekanina \& Kracht 2014).  We refer to this law as a
$g_{\rm for}(r)$ law.  We find that at a distance as close to the
Sun as 5\,{\Rsun}, $g_{\rm for}(r)$ varies as steeply as
$\sim \! r^{-18.2}$, that is, very differently from the expected
laws for water ice and sodium.

\begin{table*}[t]
\vspace{0.1cm}
\begin{center}
{\footnotesize {\bf Table 6}\\[0.1cm]
{\sc Comparison of the Momentum-Transfer Laws $g_{\rm ice}(r)$, $g_{\rm
Na}(r)$, $g_{\rm for}(r)$, and $g_{\rm mod}(r; r_0)$\\in Fitting the Orbits
of the Eight Dwarf Sungrazers from Table~4.}\\[0.15cm]
\begin{tabular}{c@{\hspace{0.55cm}}l@{\hspace{0.7cm}}l@{\hspace{0.45cm}}c@{\hspace{0.7cm}}c@{\hspace{0.55cm}}c@{\hspace{0.6cm}}c@{\hspace{0.6cm}}c@{\hspace{0.6cm}}c@{\hspace{0.5cm}}c}
\hline\hline\\[-0.25cm]
  & & Momentum & Param-
    & \multicolumn{2}{@{\hspace{-0.45cm}}c}{Distance $r_0\,^{\rm b}$}
    & \multicolumn{2}{@{\hspace{-0.4cm}}c}{Line of apsides}
    & Apsidal & RMS \\[-0.04cm]
  & & transfer & eter ${\cal A}_3$
    & \multicolumn{2}{@{\hspace{-0.45cm}}c}{\rule[0.6ex]{2.3cm}{0.4pt}}
    & \multicolumn{2}{@{\hspace{-0.4cm}}c}{\rule[0.6ex]{2.5cm}{0.4pt}}
    & line's & resid- \\[-0.04cm]
No. & Object & law & (units$^{\rm a}$) & in AU & in {\Rssun}
    & $L_\pi$ & $B_\pi$ & offset & ual \\[0.03cm]
\hline \\[-0.17cm]
1 & C/2007 X13 & $g_{\rm ice}(r)$     & \llap{$-$}0.188 & \,\,\ldots\ldots
  & \,\,\ldots\ldots & 282$^\circ\!\!$.95 & +35$^\circ\!\!$.08
  & 0$^\circ\!\!$.17 & $\pm$13$^{\prime\prime}\!\!$.0 \\
  &            & $g_{\rm Na}(r)$      & \llap{$-$}0.189 & \,\,\ldots\ldots
  & \,\,\ldots\ldots & 282.99 & +35.08 & 0.20 & $\pm$13.1 \\ 
  &            & $g_{\rm for}(r)$     & \llap{$-$}0.151 & \,\,\ldots\ldots
  & \,\,\ldots\ldots & 281.88 & +35.20 & 0.75 & $\pm$13.2 \\
  &            & $g_{\rm mod}(r; r_0)$ & \llap{$-$}0.183 & 0.126 & 27.1
  & 282.81 & +35.20 & 0.01 & $\pm$12.9 \\[0.1cm]
%
% &            & $G(r;z,\mbox{\rrend})$\rlap{$^{\rm b}$} & \llap{$-$}0.128
% & 2.15 & 0.0073\,   & 1.57\, & 282.80 & +35.20 & 0.00 & $\pm$12.8 \\[0.1cm]
%
2 & C/2007 X3  & $g_{\rm ice}(r)$     & \llap{$-$}0.111 & \,\,\ldots\ldots
  & \,\,\ldots\ldots & 284.17 & +34.44 & 1.36 & $\;\:\pm$6.6 \\
  &            & $g_{\rm Na}(r)$    & \llap{$-$}0.111 & \,\,\ldots\ldots
  & \,\,\ldots\ldots & 284.18 & +34.42 & 1.37 & $\;\:\pm$6.6 \\
  &            & $g_{\rm for}(r)$   & \llap{$-$}0.025\rlap{4} & \,\,\ldots\ldots
  & \,\,\ldots\ldots & 283.13 & +35.13 & 0.28 & $\;\:\pm$3.0 \\
  &            & $g_{\rm mod}(r;r_0)$ & \llap{$-$}0.012\rlap{1} & 0.0056 & 1.2
  & 283.08 & +35.15 & 0.23 & $\;\:\pm$2.8 \\[0.1cm]
%
% &            & $G(r;z,\mbox{\rrend}) $ & & & &
% & 281.96 & +35.10 & 0.69 % &$ \, \:\pm$3 .4 \\[0.1 cm]
%
3 & C/2001 Y4  & $g_{\rm ice}(r)$   & \llap{$-$}0.031\rlap{5} & \,\,\ldots\ldots
  & \,\,\ldots\ldots & 283.15 & +35.26 & 0.29 & $\pm$20.8 \\
  &            & $g_{\rm Na}(r)$    & \llap{$-$}0.031\rlap{2} & \,\,\ldots\ldots
  & \,\,\ldots\ldots & 283.15 & +35.24 & 0.29 & $\pm$20.8 \\
  &            & $g_{\rm for}(r)$   & \llap{$-$}0.018\rlap{6} & \,\,\ldots\ldots
  & \,\,\ldots\ldots & 283.73 & +35.21 & 0.76 & $\pm$21.0 \\
  &            & $g_{\rm mod}(r;r_0)$ & \llap{$-$}0.031\rlap{3} & 2.12 & 455
  & 283.15 & +35.23 & 0.29 & $\pm$20.8 \\[0.1 cm]
%
% &            & $G(r;z,\mbox{\rrend})$ & \\[0.1 cm]
%
% 4 & C/2009 X15 & $g_{\rm ice}(r)$ & \llap{$-$}0.025\rlap{1} & \,\,\ldots\ldots
% & \,\,\ldots\ldots & 281.10 & +35.03 & 1.40 & $\;\:\pm$9.6 \\
%
% &            & $g_{\rm Na}(r)$    & \llap{$-$}0.025\rlap{0} & \,\,\ldots\ldots
% & \,\,\ldots\ldots & 281.10 & +35.11 & 1.40 & $\;\:\pm$9.7 \\
%
% &            & $g_{\rm for}(r)$   & \llap{$-$}0.021\rlap{7} & \,\,\ldots\ldots
% & \,\,\ldots\ldots & 281.28 & +35.06 & 1.25 & $\;\:\pm$9.7 \\
%
% &            & $g_{\rm mod}(r;r_0)$ & \llap{$-$}0.016\rlap{6} & 0.0097 & 2.1
% & 281.29 & +35.08 & 1.24 & $\;\:\pm$9.6 \\[0.1cm]
%
% &            & $G(r;z,\mbox{\rrend})$ & \\[0.1cm]
%
4 & C/2008 M4  & $g_{\rm ice}(r)$   & \llap{$-$}0.022\rlap{8} & \,\,\ldots\ldots
  & \,\,\ldots\ldots & 282.43 & +35.08 & 0.33 & $\;\:\pm$8.8 \\
  &            & $g_{\rm Na}(r)$    & \llap{$-$}0.022\rlap{7} & \,\,\ldots\ldots
  & \,\,\ldots\ldots & 282.42 & +35.10 & 0.33 & $\;\:\pm$8.8 \\
  &            & $g_{\rm for}(r)$   & \llap{$-$}0.100\rlap{4} & \,\,\ldots\ldots
  & \,\,\ldots\ldots & 282.63 & +35.15 & 0.15 & $\;\:\pm$8.5 \\
  &            & $g_{\rm mod}(r;r_0)$ & \llap{$-$}0.138\rlap{8} & 0.012 & 2.6
  & 282.65 & +35.15 & 0.13 & $\;\:\pm$8.4 \\[0.1cm]
5 & C/2009 L5$^{\rm c}$ & $g_{\rm ice}(r)$ & \llap{$-$}0.003\rlap{7}
  & \,\,\ldots\ldots & \,\,\ldots\ldots & 282.37 & +35.28 & 0.36
  & $\;\:\pm$6.9 \\
  &              & $g_{\rm Na}(r)$  & \llap{$-$}0.003\rlap{6} & \,\,\ldots\ldots
  & \,\,\ldots\ldots & 282.37 & +35.28 & 0.36 & $\;\:\pm$6.9 \\
  &              & $g_{\rm for}(r)$ & \llap{$-$}0.004\rlap{6} & \,\,\ldots\ldots
  & \,\,\ldots\ldots & 282.37 & +35.29 & 0.36 & $\;\:\pm$6.6 \\[0.1cm]
%
% &              & $g_{\rm mod}(r; r_0)$   & & & & & & & \\[0.1cm]
%
% &              & $G(r;z,\mbox{\rrend})$ & \\[0.1cm]
%
6 & C/2006 J9    & $g_{\rm ice}(r)$   & \llap{+}0.006\rlap{2} & \,\,\ldots\ldots
  & \,\,\ldots\ldots & 282.39 & +35.88 & 0.76 & $\;\:\pm$3.9 \\
  &              & $g_{\rm Na}(r)$    & \llap{+}0.006\rlap{2} & \,\,\ldots\ldots
  & \,\,\ldots\ldots & 282.39 & +35.88 & 0.76 & $\;\:\pm$3.9 \\
  &              & $g_{\rm for}(r)$   & \llap{+}0.002\rlap{4} & \,\,\ldots\ldots
  & \,\,\ldots\ldots & 283.48 & +36.24 & 1.18 & $\;\:\pm$3.9 \\
  &              & $g_{\rm mod}(r; r_0)$ & \llap{+}0.006\rlap{7} & 0.052
  & 11.2 & 282.44 & +35.78 & 0.65 & $\;\:\pm$3.9 \\[0.1cm]
%
% &              & $G(r;z,\mbox{\rrend})$ & \\[0.1cm]
%
7 & C/2008 M5$^{\rm d}$ & $g_{\rm ice}(r)$   & \llap{+}0.028\rlap{8}
  & \,\,\ldots\ldots & \,\,\ldots\ldots & 282.67 & +35.06 & 0.18
  & $\;\:\pm$9.2 \\
  &              & $g_{\rm Na}(r)$    & \llap{+}0.028\rlap{7} & \,\,\ldots\ldots
  & \,\,\ldots\ldots & 282.67 & +35.06 & 0.17 & $\;\:\pm$9.2 \\
  &              & $g_{\rm for}(r)$   & \llap{+}0.074\rlap{5} & \,\,\ldots\ldots
  & \,\,\ldots\ldots & 281.96 & +35.05 & 0.70 & $\;\:\pm$9.5 \\[0.1cm]
%
% &              & $g_{\rm mod}(r; r_0)$   & & & & & & & \\[0.1cm]
%
% &              & $G(r;z,\mbox{\rrend})$ & \\[0.1cm]
%
8 & C/2008 K8\rlap{$^{\rm e}$}  & $g_{\rm ice}(r)$   & \llap{+}0.024\rlap{3}
  & \,\,\ldots\ldots & \,\,\ldots\ldots & 282.38 & +35.36 & 0.38 & $\pm$10.3 \\
  &              & $g_{\rm Na}(r)$ & \llap{+}0.024\rlap{2} & \,\,\ldots\ldots
  & \,\,\ldots\ldots & 282.39 & +35.38 & 0.38 & $\pm$10.3 \\
  &              & $g_{\rm mod}(r; r_0)$ & \llap{+}0.028\rlap{4} & 0.080 & 17.2
  & 282.49 & +35.33 & 0.29 & $\pm$10.4 \\[0.1cm]
%
% &              & $g_{\rm for}(r)$ & & & \llap{no solution}$\!$
% & & & & \\
%
% &              & $g_{\rm mod}(r; r_0)$ & \llap{+}0.037\rlap{3} & 0.065 & 14.0
% & 282.56 & +35.30 & 0.21 & $\pm$10.5 \\[0.1cm]
%
% &              & $G(r;z,\mbox{\rrend})$ & \\[0.08cm]
%
\hline \\[-0.23cm]
\multicolumn{10}{l}{\parbox{15.23cm}{$^{\rm a}$\,{\scriptsize Normal
 component of the momentum-transfer acceleration at a heliocentric distance
 {\vspace{-0.1cm}}of 10\,{\Rssun}, expressed in AU day$^{-2}$.~The  Sun's
 gravitational acceleration at this distance is 0.1366 AU
 day$^{-2}$.}}}\\[0.12cm]
\multicolumn{10}{l}{\parbox{15.23cm}{$^{\rm b}$\,{\scriptsize By definition,
 scaling distance $r_0$ is always equal to \mbox{2.808 AU = 603 {\Rssun}}
 for law{\vspace{-0.08cm}} $g_{\rm ice}(r)$; \mbox{0.3458 AU = 74.3 {\Rssun}}
 for $g_{\rm Na}(r)$; and \mbox{0.01486 AU = 3.19 {\Rssun}} for $g_{\rm
 for}(r)$ [Sekanina \& Kracht 2014].}}}\\[0.06cm]
\multicolumn{10}{l}{\parbox{15.23cm}{$^{\rm c}$\,{\scriptsize There was no
 minimum offset from the reference apsidal line among modified solutions
 {\vspace{-0.07cm}}with $r_0$ between 0.02 and~4~AU, although the RMS residual
 was decreasing steadily with{\vspace{-0.07cm}} decreasing $r_0$, which is
 consistent with the lower RMS residual from the solution with the
 forsterite-based sublimation law.}}}\\[-0.02cm]
\multicolumn{10}{l}{\parbox{15.23cm}{$^{\rm d}$\,{\scriptsize There was no
 minimum offset from the reference apsidal line among modified solutions
 {\vspace{-0.05cm}}with $r_0$ between 0.02 and 2.8~AU.}}}\\[0.06cm]
\multicolumn{10}{l}{\parbox{15.23cm}{$^{\rm e}$\,{\scriptsize The solution
 optimized for this object with law $g_{\rm for}(r)$ resulted in
 \mbox{$A_3 = 0$}, i.e., in a gravitational solution (Table
 4).}{\vspace{0.08cm}}}}
%
% \multicolumn{10}{l}{\parbox{15.1cm}{$^{\rm b}$\,{\scriptsize There is
% a set of equally satisfactory solutions (in terms of the RMS error) that
% extend{\vspace{-0.05cm}} along a trough from 1.51 to $\sim$2.4 in $z$ and
% from 3.7\,{\Rssun} to 1\,{\Rssun} in {\rrend}.}}}
%
\end{tabular}}
\end{center}
\end{table*}

The third law introduced in this section is a generic one, which we
refer to as a {\it modified nongravitational law\/} $g_{\rm mod}(r;r_0)$
[again normalized to \mbox{$g_{\rm mod}(1\;{\rm AU}; r_0) = 1$}] and which
is aimed at obtaining the best possible fit by the law of type (22) to the
astrometric observations by varying {\it only\/} the scaling distance $r_0$.
This approach is justified by
Marsden et al.'s (1973) finding that the shapes of normalized sublimation
curves for a variety of species are fairly similar except for major
horizontal shifts in a plot of log\,(sublimation rate) against $\log r$,
which means that in terms of the approximation formula (22) the curves
are relatively insensitive to the exponents $m$ (which always slightly
exceeds 2), $n$, and $k$, but highly sensitive to the scaling distance
$r_0$.  On some assumptions (see Sekanina \& Kracht 2014), $r_0$
measures essentially the heat of sublimation $L$ of the outgassing
substance (or a mean value, if more species are involved), varying to a
first approximation inversely as the square of $L$,
\begin{equation}
r_0 \simeq \left( \frac{\rm const}{L} \right)^{\!2} \!,
\end{equation}
where a calibration by water ice gives{\vspace{-0.06cm}} for the{\nopagebreak}
constant 19\,100~AU$^{\frac{1}{2}}$~cal~mol$^{-1}$ in the case of an isothermal
model.

To test the modified nongravitational law on the eight dwarf Kreutz
sungrazers, we retain the values of $m$, $n$, and $k$ for water ice
from Eq.\,(22), but conduct in each case a search for an optimum
solution by varying the scaling distance $r_0$ until a minimum offset
from the reference apsidal orientation is found, as described in
Sec.~5.2.
%
% The $g_{\rm ice}(r)$ and $g_{\rm for}(r)$ laws are in the range of
% heliocentric distances 3--30\,{\Rsun} compared in Figure~8.  The
% deviations from a simple power law are seen to be relatively minor
% (for forsterite) to completely negligible (for water ice).
%
%
% \begin{figure}
% \vspace{-3.67cm}
% \hspace{0.17cm}
% \centerline{
% \scalebox{0.68}{
% \includegraphics{f8_SOHOkreutz.ps}}} % from f8_SOHOkreutz.tex   OUT
% 
% \vspace{-4.85cm}
% \caption{Comparison of the momentum-transfer effects:\ for water
% ice, $g_{\rm ice}(r)$; forsterite, $g_{\rm for}(r)$; and generic curves,
% $G(r;z,\mbox{\rrend})$, which include the effect of a dwindling size of
% the nucleus.  The generic curve for $z = 10$, split near the lower end
% of heliocentric distances, shows the role of {\rrend}:\ a higher
% momentum-transfer effect refers to a greater disintegration distance.
% The accelerations are arbitrarily normalized to a distance of
% 10\,{\Rsun}.{\vspace{0.3cm}}}
% \end{figure}

The optimum nongravitational solutions, derived with the four different
nongravitational laws, $g_{\rm ice}(r)$, $g_{\rm Na}(r)$, $g_{\rm for}(r)$,
and $g_{\rm mod}(r; r_0)$, are compared in Table~6.  The data for the
$g_{\rm ice}(r)$ law are taken over from the ($A_3$)~rows of Table~4,
except that the nongravitational parameter in the direction normal to
the orbit plane, now denoted ${\cal A}_3$, is referred to a~heliocentric
{\vspace{-0.06cm}}distance of 10\,{\Rsun} (and~ex\-pressed in AU day$^{-2}$).

Inspection of Table 6 suggests the following:\ (i)~the resulting values
of ${\cal A}_3$ from the solutions based on the standard law and the sodium
sublimation law are practically identical, as expected, for all eight test
sungrazers; (ii)~the nongravitational acceleration for C/2007~X13 at 10~{\Rsun}
from the Sun exceeds the Sun's gravitational acceleration by up to nearly
40~percent; for the other seven comets it is smaller, but it always amounts to
more than 1~percent of the Sun's attraction; (iii)~for five comets (C/2007~X13,
C/2007~X3, C/2008~M4, C/2006~J9, and C/2008~K8) the modified nongravitational
law provides the best solution in terms of the apsidal-line offset; while for
C/2001~Y4 the solutions based on the modified law, the standard law, and the
sodium sublimation law all fit the data equally well; no minimum apsidal-line
offset was found among the solutions based on the modified law for C/2009~L5
and C/2008~M5; (iv)~for C/2009~L5 the ice, sodium, and forsterite sublimation
laws offer equally good solutions in terms of the apsidal-line offset, but
the solution based on the forsterite sublimation law provides a~fit with
a RMS residual superior to those from the solutions based on the other two
laws; (v)~for C/2008~M5, the best fit results by a narrow margin from the
solution based on the sodium sublimation law; (vi)~the forsterite sublimation
law does not work for C/2008~K8; (vii)~in terms of the scaling distance of the
modified law, the comets are divided into three groups:\ two (C/2007~X3
and C/2008~M4) have the nongravitational variations steeper than even the
forsterite sublimation law; three (C/2006~J9, C/2008~K8, and C/2007~X13) have
the variations steeper than the sodium sublimation law, but less steep than
the forsterite sublimation law; and only one (C/2001~Y4) has variations less
steep than the sodium law, though still steeper than the ice sublimation law.

To summarize, in terms of the apsidal-line offset and the RMS residual, the
water-ice sublimation law proves competitive with the other three laws only
in the case of C/2001~Y4.  For the remaining seven comets, the laws with
smaller scaling distances, that is, implying steeper variations, are superior,
suggesting that momentum-transfer effects driven by the sublimation of
species substantially less volatile than water ice dominate the motions of
these Kreutz sungrazers.

\section{Three-Parameter Nongravitational\\Solutions for Dwarf Kreutz
Sungrazers}
In Sec.~5.2 we already remarked that among comets with perihelion distances
of \gapeq$\!$1~AU that required an incorporation of the nongravitational terms
into the equations of motion, almost never was there the need to include
a normal component, $A_3$.  One of very few exceptions was the case of
comet 71P/Clark, for which inclusion of $A_3$ was necessary in order to
link the apparitions 1995--2000 and 1995--2006 (Nakano 2001, 2006, 2008). 

From what we have until now determined in this paper, the dwarf sungrazers
of the Kreutz system are a major exception to the rule of \mbox{$A_3
\rightarrow 0$}:\ the normal component always appears to play a role in
their orbital motions.  However, since we have not up to this point
investigated the contributions from $A_1$ and $A_2$ (both of which having
been assumed zero), it is unclear whether or not the normal component
actually {\it dominates\/} the other two in magnitude.

Since the nongravitational parameters $A_1$, $A_2$, and $A_3$ cannot
satisfactorily be determined in the course of optimizing an orbital
solution by directly incoporating them as parametric variables into the
equations of motion when fitting the astrometric observations, it is
necessary to explore this issue by employing iterations.

We begin with the relations (3) between the instantaneous rates of change
in the angular orbital elements, $d\omega/dt$, $d\Omega/dt$, and $di/dt$
on the one hand and the acceleration components, which we now express as
a function of a dimensionless, normalized nongravitational law $g(r)$, on
the other hand.  We note that $g(r)$ stands for any of the $g_{\rm ice}(r)$,
\ldots, $g_{\rm mod}(r; r_0)$ laws:
\begin{equation}
\left( \!
\begin{array}{c}
j_{\rm R}(t) \\
j_{\rm T}(t) \\
j_{\rm N}(t) \end{array}
\! \right) = \left( \!\!
\begin{array}{c}
A_1 \\ A_2 \\ A_3 \end{array}
\!\! \right) \cdot g(r).
\end{equation}
Integrating over the period of observation, from $t_{\rm beg}$ to $t_{\rm
fin}$, we find for the overall increments in the three angular elements:
\begin{equation}
\left(\!\! \begin{array}{c}
\Delta \omega \\ \Delta \Omega \\ \Delta i
\end{array}
\!\! \right) = \left( \! \begin{array}{ccc}
\Im_{11} & \Im_{21} & \Im_{31} \\
0\, & 0\, & \Im_{32} \\
0\, & 0\, & \Im_{33}
\end{array} \!\! \right) \!\cdot\! \left( \!\! \begin{array}{c}
A_1 \\ A_2 \\ A_3
\end{array}
\!\! \right) \! ,
\end{equation}
where
\begin{eqnarray}
\Im_{11} & = & \!\int_{t_{\rm beg}}^{t_{\rm fin}} \!\!\!X_{\rm R}\,g(r)\,dt,
 \nonumber \\
\Im_{21} & = & \!\int_{t_{\rm beg}}^{t_{\rm fin}} \!\!\!X_{\rm T}\,g(r)\,dt,
 \nonumber \\
\Im_{31} & = & \!\int_{t_{\rm beg}}^{t_{\rm fin}} \!\!\!X_{\rm N}\,g(r)\,dt, \\ 
\Im_{32} & = & \!\int_{t_{\rm beg}}^{t_{\rm fin}} \!\!\!Y_{\rm N}\,g(r)\,dt,
 \nonumber \\
\Im_{33} & = & \!\int_{t_{\rm beg}}^{t_{\rm fin}} \!\!\!Z_{\rm N}\,g(r)\,dt,
 \nonumber
\end{eqnarray}
and $r = r(t)$.  Inserting Eqs.\ (25) and (26) into the first of Eqs.\ (9),
the established constraint \mbox{$\Delta L_\pi = 0$} offers the following
relationship among $A_1$, $A_2$, and $A_3$:
\begin{equation}
A_1 \Im_{11} + A_2 \Im_{21}
             + A_3 \Im\mbox{\large \boldmath $^\ast$}_{\:\!\!\!\!\!31} = 0,
\end{equation}
where
\begin{equation}
\Im\mbox{\large \boldmath $^\ast$}_{\:\!\!\!\!\!31} = \Im_{31} + 
  \Im_{32} \:\! \frac{1 \!-\! \sin^2 \omega \sin^2 i}{\cos i}
  \:\!-\:\! {\textstyle \frac{1}{2}} \Im_{33} \sin 2\omega \tan i.\!
\end{equation}

In reality, the constraint \mbox{$\Delta L_\pi = 0$} is of course
valid only statistically, as Figure~3 and Tables~2 and 3 show.  Since
\mbox{$A_3 \neq 0$} and since the type of solutions investigated up to
now have been based on a constraint \mbox{$A_1 \!=\! A_2 \!=\! 0$},
condition (27) inevitably requires that \mbox{$\Im\mbox{\large \boldmath
$^\ast$}_{\:\!\!\!\!\!31} \simeq 0$}, in which case
\begin{equation}
A_1 \Im_{11} + A_2 \Im_{21} \simeq 0.
\end{equation}
This relation shows that there is no reason why $A_1$ and $A_2$ should be
zero; in fact, there is an infinite number of nonzero $(A_1, A_2)$
pairs that satisfy the condition (29).  Some of the pairs, in which
\mbox{$A_1 \neq 0$} and \mbox{$A_2 \neq 0$}, may provide even
a better match to the reference apsidal line orientation than does the case
\mbox{$A_1 \!=\! A_2 \!=\! 0$}.  To determine what pair of $A_1$ and $A_2$
offers --- on the statistically valid condition of \mbox{$\Delta L_\pi = 0$}
--- a solution optimized in terms of a minimum apsidal offset is the final
objective of this investigation.

\begin{table*}[t]
\vspace{0.1cm}
\begin{center}
{\footnotesize {\bf Table 7}\\[0.1cm]
{\sc Final Nongravitational Orbital Elements for Eight Test Dwarf Kreutz
Sungrazers (Condition \mbox{$\Delta L_\pi = 0$}; Equinox J2000).}\\[0.1cm]
\begin{tabular}{@{\hspace{0.03cm}}c@{\hspace{0.08cm}}l@{\hspace{0.14cm}}c@{\hspace{0.17cm}}c@{\hspace{0.05cm}}c@{\hspace{0.12cm}}c@{\hspace{0.13cm}}c@{\hspace{0.18cm}}c@{\hspace{0.11cm}}c@{\hspace{0.18cm}}c@{\hspace{0.13cm}}c@{\hspace{0.13cm}}c@{\hspace{0.23cm}}c@{\hspace{-0.05cm}}c@{\hspace{-0.07cm}}c@{\hspace{0.02cm}}c@{\hspace{0.04cm}}}
\hline\hline\\[-0.2cm]
 & & \multicolumn{5}{@{\hspace{0cm}}c}{Orbital elements$^{\rm a}$}
   & \multicolumn{5}{@{\hspace{0cm}}c}{Nongravitational law and
     parameters$^{\rm b}$}
   & \multicolumn{2}{@{\hspace{0.05cm}}c}{Line\,of\,apsides} & Apsidal
   & \,RMS \\[-0.04cm]
 & & \multicolumn{5}{@{\hspace{0cm}}c}{\rule[0.6ex]{6.05cm}{0.4pt}}
   & \multicolumn{5}{@{\hspace{-0.02cm}}c}{\rule[0.6ex]{5.1cm}{0.4pt}}
   & \multicolumn{2}{@{\hspace{0.05cm}}c}{\rule[0.6ex]{1.9cm}{0.4pt}}
   & line's & \,resid- \\[-0.04cm]
No. & \hspace{0.35cm}Object & $t_\pi$\,(ET) & $\omega$ & $\Omega$ & $i$
    & \,$q$ & law & $[\:\!r_0\:\!]^{\rm c}$ & ${\cal A}_1$ & ${\cal A}_2$
    & ${\cal A}_3$ & $L_\pi$ & $B_\pi$ & offset & \,ual \\[0.08cm]
\hline \\[-0.2cm]
1 & C/2007 X13 & 2007/12/14.523 & 87$^\circ\!\!$.349
  & $\;\:\;\:$9$^\circ\!\!$.567 & 144$^\circ\!\!$.753 & 1.023 & MD
  & 0.126 & \,\,\ldots\ldots & \,\,\ldots\ldots & $-$0.1831
  & 282$^\circ\!\!$.81 & +35$^\circ\!\!$.20 & 0$^\circ\!\!$.01
  & $\pm$12$^{\prime\prime}\!\!$.9 \\
%
% 2 & C/2007 X3 & 2007/12/05.219 & 79.954 & $\;\:\;\:$0.763 & 144.220 & 1.225
%   & MD & 0.0056 & \,\,\ldots\ldots & \,\,\ldots\ldots & $-$0.0121
%   & 283.08 & +35.15 & 0.23 & $\;\:\pm$2.8 \\
%
2 & C/2007 X3 & 2007/12/05.219 & 80.291 & $\;\:\;\:$0.893 & 144.210 & 1.211
  & MD & 0.0056 & \llap{(}$-$0.0265 & +0.0055 & +0.0065\rlap{)} & 282.80
  & +35.20 & 0.00 & $\;\:\pm$2.3 \\
3 & C/2001 Y4 & 2001/12/18.662 & 83.412 & $\;\:\;\:$4.961 & 144.505 & 1.020
  & MD & 2.12 & +0.0808 & $-$0.0124 & $-$0.0419 & 283.04 & +35.23 & 0.19
  & $\pm$21.8 \\
4 & C/2008 M4 & 2008/06/25.713 & 78.839 & 359.095 & 144.018 & 1.013 & MD
  & 0.012 & \llap{(}$-$0.1027 & +0.0165 & $-$0.0829\rlap{)} & 282.80 & +35.20
  & 0.00 & $\;\:\pm$8.3 \\
5 & C/2009 L5 & 2009/06/05.330 & 78.553 & 358.534 & 143.901 & 1.031 & FT
  & \,\,\ldots\ldots & +0.0557 & $-$0.0099 & $-$0.0418 & 282.60 & +35.27 & 0.18
  & $\;\:\pm$6.2 \\
6 & C/2006 J9 & 2006/05/10.964 & 83.884 & $\;\:\;\:$5.177 & 144.431 & 1.058
  & MD & 0.052 & +0.0246 & $-$0.0044 & +0.0069 & 282.68 & +35.34
  & 0.17 & $\;\:\pm$4.1 \\
7 & C/2008 M5 & 2008/06/26.371 & 82.383 & $\;\:\;\:$3.356 & 144.582 & 1.082
  & NA & \,\,\ldots\ldots & \,\,\ldots\ldots & \,\,\ldots\ldots & +0.0287
  & 282.67 & +35.06 & 0.17 & $\;\:\pm$9.2 \\
8 & C/2008 K8 & 2008/05/28.534 & 83.630 & $\;\:\;\:$4.996 & 144.562 & 1.032
  & MD & 0.080 & +0.0376 & $-$0.0060 & +0.0110 & 282.80 & +35.19 & 0.01
  & $\pm$10.2 \\[0.05cm]
\hline \\[-0.23cm]
\multicolumn{16}{l}{\parbox{17.6cm}{$^{\rm a}$\,{\scriptsize Perihelion
 distance $q$ is expressed in units of solar radii, {\Rssun}
 (1\,{\Rssun}\,=\,0.0046548 AU).}}}\\[-0.04cm]
\multicolumn{16}{l}{\parbox{17.6cm}{$^{\rm b}$\,{\scriptsize MD\,=\,modified
 law; FT\,=\,forsterite sublimation law ($r_0\!=\!0.01486$\,AU); NA\,=\,sodium
 {\vspace{-0.04cm}}sublimation law ($r_0\!=\!0.3458$\,AU); scaling distance
 of {\hspace*{0.22cm}}fitted modified law $r_0$ is expressed in AU; the
 acceleration components {\vspace{-0.08cm}}${\cal A}_1$, ${\cal A}_2$, and
 ${\cal A}_3$ are referred to heliocentric distance of 10\,{\Rssun} and
 are {\hspace*{0.21cm}}expressed in units of AU\,day$^{-2}\!$.  Sun's
 gravitational acceleration at 10\,{\Rssun} is 0.1366 AU day$^{-2}$ or
 273.7\,cm\,s$^{-2}$.}}}\\[0.3cm]
\multicolumn{16}{l}{\parbox{17.6cm}{$^{\rm c}$\,{\scriptsize Distance $r_0$
 is in AU; in units of {\Rssun}, the values are, from top to bottom:\ 27.1,
 1.2, 455, 2.6, 11.2, and 17.2.  The {\vspace{-0.04cm}}range of
 \mbox{heliocentric distances} {\hspace*{0.2cm}}spanned by the observations
 {\vspace{-0.04cm}}with the C2 coronagraph is:\ 0.0587--0.0461~AU (or
 12.6--9.9~{\Rssun}) for C/2007~X13; 0.0551--0.0375~AU (or 11.8--8.1~{\Rssun})
 {\hspace*{0.2cm}}for C/2007~X3; 0.0569--0.0420~AU (or 12.2--9.0~{\Rssun})
 for C/2001~Y4;{\vspace{-0.04cm}} 0.0581--0.0441~AU (or 12.5--9.5~{\Rssun}) for C/2008~M4;
 0.0588--0.0385~AU {\hspace*{0.18cm}}(or 12.6--8.3~{\Rssun}) for C/2009~L5;
 0.0552--0.0375~AU (or 11.9--8.1~{\Rssun}) for C/2006~J9; 0.0586--0.0429~AU
 (or 12,6--9.2~{\Rssun}) for C/2008~M5;{\vspace{-0.04cm}} and
 {\hspace*{0.2cm}}0.0675--0.0514~AU (or 14.5--11.0~{\Rssun}) for C/2008~K8.}}}
\end{tabular}}
\end{center}
\end{table*}

The contributions to the orbital solution from the parameter $A_3$ on the
one hand and from $A_1$ and $A_2$ on the other hand are now separated from
each other, and the extension of our work --- a transition from solutions
with $A_3$ to those with all three parameters --- is accomplished with the
aid of Eq.\,(29).  Keeping $A_3$ constant and equivalent to ${\cal A}_3$
in Table~6, we continue to search for a pair of $A_1$ and $A_2$ such that
it results in a minimum apsidal-line offset; we successively iterate $A_1$
and find $A_2$ from
\begin{equation}
A_2 \simeq - \frac{\Im_{11}}{\Im_{21}} A_1,
\end{equation}
with $\Im_{11}$ and $\Im_{21}$ computed from Eqs.\,(26) by numerical
integration of the expressions, in which the $g(r)$ again stands for
any of the employed nongravitational laws.  Once a minimum offset from
the apsidal line is found for an adopted $A_3$, the resulting $A_1$ and
$A_2$ are kept constant and a search for a new $A_3$ initiated by further
optimizing the apsidal line, etc., until the offset's ultimate minimum
is found.  In practice, the momentum-transfer law selected for this
approach should be the one providing the least offset from the reference
apsidal orientation in Table~6, which, as it turns out, is in most cases
the modified law $g_{\rm mod}$.

The results of these computations are presented in Table 7, which shows
that our final sets of orbital elements for all eight test sungrazers
match the reference direction of the line of apsides to within
0$^\circ\!$.2, which is --- as seen from Table~1 --- its intrinsic
uncertainty.  For two comets in Table~7, C/2007~X13 and C/2008~M5, no
three-parameter nongravitational solution has been attempted, because
a single-parameter solution already implies an offset smaller than the
stipulated limit of 0$^\circ\!$.2.  For the remaining entries of Table~7,
the three-parameter solution was successfully carried out, yielding in
most cases the radial component positive and dominating the other two
components.  The exceptions are C/2007~X3 and C/2008~M4, for which the
radial component came out to be negative, for which we do not have an
explanation.

The sungrazers C/2007 X3 and C/2008 M4 are, together with C/2009~L5, the
objects in Tables 6 and 7 with the most steeply varying nongravitational
accelerations, characterized by the smallest scaling distances $r_0$.  Next
comes another group of three --- C/2006~J9, C/2007~X13, and C/2008~K8 ---
whose scaling distances are in a range from 0.05~AU to 0.15~AU, intermediate
between those of the forsterite sublimation law and the sodium sublimation
law.  The equivalent sublimation heat is estimated at 50\,000 to 80,000 cal
mol$^{-1}$.  The motion of C/2008~M5 was fitted best by the sodium sublimation
law, and the volatile end is represented by C/2001~Y4, whose motion appears to
have been affected by sublimation of species that may have included water ice;
the effective sublimation heat is estimated at between 13\,000 and 15\,000~cal
mol$^{-1}$.  Because the laws applicable to the last two sungrazers vary
essentially as an inverse square of heliocentric distance along the observed
arcs of the orbits, there could be contributions from the solar radiation
pressure (but only in ${\cal A}_1$ of course), if the nuclei of these comets
were already shattered into dust at the time.

The very high nongravitational accelerations, already mentioned in Sec.~5.2,
are fully confirmed, including the record value of ${\cal A}_3$ for C/2007~X13,
which is 134~percent of the Sun's gravitational acceleration.  Similarly, the
overall nongravitational accelerations of C/2001~Y4 and C/2009~L5 amount to,
respectively, 67 and 51~percent of the Sun's attraction.  Even the least
overall nongravitational accelerations in Table~7 are still on the order
of $\sim$20~percent of the Sun's attraction.

There are at least three physical processes that operate on the dwarf
sungrazers in the final phase of their disintegration near the end of the
visible trajectory and are responsible for the very high nongravitational
accelerations applied to the test objects in Table 7:\ sublimation,
fragmentation, and the Sun's radiation pressure.  The joint contribution to
the acceleration from the first two processes, which result in erosion of the
sungrazer's nucleus, with a progressive loss rate of its mass, can numerically
be simulated by the law $g(r)$, whereas the radiation pressure acceleration
varies of course as $r^{-2}$ except at close proximity to the Sun, where
it increases as \mbox{$2 [1 \!-\!  \sqrt{1 \!-\!  (\mbox{\Rssun}/r)^2}]$};
this rate of variation is 1.03 times steeper than the inverse square law
3\,{\Rsun} from the Sun, 1.15 times steeper at 1.5\,{\Rsun}, 1.29 times
at 1.2\,{\Rsun}, and 1.53 times steeper at 1.05\,{\Rsun}.

We begin with{\vspace{-0.05cm}} the conservation of momentum law, which
requires that a relative mass erosion rate, $\dot{\cal M}$, of a comet
generates an acceleration $\gamma$ on its nucleus of mass ${\cal M}$,
which at time $t$ satisfies a relation
\begin{equation}
\gamma(t) {\cal M}(t) = -\upsilon(t) \kappa(t) \dot{\cal M}(t),
\end{equation}
where $\upsilon(t)$ is the outflow velocity of the eroded mass,
$\kappa(t)$ is a factor that accounts for the degree of its collimation,
and the minus sign indicates that $\upsilon(t)$ and $\gamma(t)$, both
taken here as positive quantities, point in opposite directions.  The
acceleration is in the following written in terms of its magnitude at
\mbox{$r_{_{\mbox{\boldmath $\!\ast$}}}\!=\!r(t_{_{\mbox{\boldmath
$\ast$}}}) \!= \! 10$\,{\Rsun}},
\begin{equation}
\gamma(t) = {\cal A}_{_{\mbox{\boldmath $\ast$}}} \,
 \frac{g(r)}{g(r_{_{\mbox{\boldmath $\!\ast$}}})} ,
\end{equation}
where \mbox{$r = r(t)$} and \mbox{${\cal A}_{_{\mbox{\boldmath $\ast$}}} =
{\cal A}(r_{_{\mbox{\boldmath $\!\ast$}}}) = \sqrt{{\cal A}_1^2 + {\cal A}_2^2
+ {\cal A}_3^2}$} from columns 10 to 12 of Table~7, while the nongravitational
law, from column~8, is at the heliocentric distances $r$ and{\vspace{-0.04cm}}
$r_{_{\mbox{\boldmath $\!\ast$}}}$ equal to, respectively, $g(r)$ and
$g(r_{_{\mbox{\boldmath $\!\ast$}}})$.  Because the observed orbital arcs
of the test sungrazers are very short, we conveniently approximate the
nongravitational law by a ``local'' power law,
\begin{equation}
\gamma(t) = {\cal A}_{_{\mbox{\boldmath $\ast$}}} \! \left(\!
 \frac{\,r_{_{\mbox{\boldmath $\!\ast$}}}}{r} \!\right)^{\!\zeta} \!\! ,
\end{equation}
where an effective exponent $\zeta(r)$ near $r(t)$ is related to the
parameters of the $g(r)$-type law from Eq.\,(22) by
\begin{equation}
\zeta(r) = m \!+\! \frac{nk}{1 \!+\! (r_0/r)^n}.
\end{equation}
Next, we approximate $\upsilon(t)$ by a thermal{\vspace{-0.09cm}}
velocity, which equals to \mbox{$\upsilon_{\rm th}(r) = \upsilon_0 (1\,{\rm
AU}/r)^{\frac{1}{4}}$}.  At 1~AU {\vspace{-0.04cm}}from the Sun, we find
$\upsilon_0$ equal to 0.44~km~s$^{-1}$ for forsterite, 0.78~km~s$^{-1}$ for
sodium atoms, and 0.54~km~s$^{-1}$ for water ice.  Averaging, we accept
\mbox{$\bar{\upsilon}_0 \approx 0.6$ km s$^{-1}$}.

The range of possible values for the collimation factor is \mbox{$0 \leq
\kappa \leq 1$}.  Having no clue for preferring any particular value, we
adopt, conditionally, \mbox{$\bar{\kappa} \approx 0.5$}.

With the help of Eq.\,(33) and these approximations, we now write Eq.\,(31)
after integration over an interval of observations, from $t_{\rm beg}$ to 
$t_{\rm fin}$,
\begin{equation}
\log_{\:\!\rm e} \! \frac{{\cal M}(t_{\rm fin})}{{\cal M}(t_{\rm beg})} =
 - \frac{{\cal A}_{_{\mbox{\boldmath $\ast$}}} r_{_{\mbox{\boldmath
 $\!\ast$}}}^{\:\!\zeta}}{\bar{\upsilon}_0 \bar{\kappa}} \, (1\,{\rm
 AU})^{-\frac{1}{4}} \!\!\!  \int_{t_{\rm beg}}^{t_{\rm fin}} \!\!
 r^{\frac{1}{4}-\zeta} dt.
\end{equation}
The integral on the right-hand side can be solved by replacing time $t$ with
a dimensionless variable, involving the perihelion distance $q$ and a
heliocentric distance $r(t)$,
\begin{equation}
x(t) = \frac{q}{r(t)}.
\end{equation}
An interval between a preperihelion time $t$ and the perihelion time $t_\pi$
is for a parabolic motion related to $x$ by
\begin{equation}
t \!-\! t_\pi = -c_0 q^{\frac{3}{2}} x^{-\frac{3}{2}} (1 \!+\! 2x)
 (1 \!-\! x)^{\frac{1}{2}},
\end{equation}
with $c_0 \!=\! 27.38$ day\,AU$^{-\frac{3}{2}}$.  Differentiating Eq.\,(37)
and inserting $dx$ for $dt$ in Eq.\,(35), we find
\begin{equation}
\log_{\:\!\rm e} \! \frac{{\cal M}(t_{\rm fin})}{{\cal M}(t_{\rm beg})}
 \!=\!  - \frac{c_1 q^{\frac{7}{4}} {\cal A}_{_{\mbox{\boldmath
 $\ast$}}}}{\bar{\upsilon}_0 \bar{\kappa}} \! \left( \!
 \frac{\,r_{_{\mbox{\boldmath $\!\ast$}}}}{q} \!  \right)^{\!\!\zeta}
 \!\! \int_{x_{\rm beg}}^{x_{\rm fin}} \!\!\! x^{\zeta - \frac{11}{4}}
 (1 \!-\! x)^{-\frac{1}{2}} dx,
\end{equation}
where{\vspace{-0.05cm}} \mbox{$c_1 \!=\! \frac{3}{2}c_0 (1\,{\rm
AU})^{-\frac{1}{4}} \!= 41.07$\,day\,AU$^{-\frac{7}{4}}$}, $\bar{\upsilon}_0$
is in AU day$^{-1}$, and $x_{\rm beg}$ and $x_{\rm fin}$ satisfy Eq.\,(36).
Writing the integral in terms of the incomplete beta function and expressing
$\bar{\upsilon}_0$ in km~s$^{-1}$, the solution becomes
\begin{equation}
\frac{{\cal M}_{\rm fin}}{{\cal M}_{\rm beg}} \!=\! \exp \left\{
 \frac{C {\cal A}_{_{\mbox{\boldmath $\ast$}}} r_{_{\mbox{\boldmath
 $\!\ast$}}}^{\:\!\zeta}}{\bar{\upsilon}_0 \bar{\kappa} q^{\,\zeta \!-\!
 \frac{7}{4}}} \! \left[ \:\!\! \raisebox{1.5ex}[1.2ex][1ex]{}
 B_{x_{\rm beg}} \:\!\!\!\!\left( \zeta \!-\!  {\textstyle \frac{7}{4}}, \!
 {\textstyle \frac{1}{2}} \right) \!-\! B_{x_{\rm fin}} \!\!
 \left( \zeta \:\!\!\!-\!\!\!\: {\textstyle \frac{7}{4}}, \! {\textstyle
 \frac{1}{2}} \right)  \right] \! \right\} \!\:\! ,
\end{equation}
where \mbox{$C\!=\!7.11\!\times\!10^4\:\!$km\,s$^{-1}\!$AU$^{-\!\frac{11}{4}}
$\,day$^{\:\!2}\:\!\!$},\,\mbox{${\cal M}_{\rm beg} \!=\! {\cal M}(t_{\rm
beg})$}, \mbox{${\cal M}_{\rm fin} = {\cal M}(t_{\rm fin})$}, and the
incomplete beta function is
\begin{equation}
B_y(\mu,\nu) = \!\! \int_{0}^{y} \!\! z^{\mu-1} (1 \!-\! z)^{\nu-1} dz,
\end{equation}
where \mbox{$\mu > 0$}, \mbox{$\nu > 0$}, and \mbox{$0 < y < 1$}.  This
definition requires a condition \mbox{$\zeta > \frac{7}{4}$}, which is
always satisfied because \mbox{$m > 2$} and the second term on the
right-hand side of Eq.\,(34) is positive.

We evaluated the exponential in Eq.\,(39) for four of the test comets in
Table~7 (C/2001~Y4, C/2009~L5, C/2006~J9, and C/2008~K8).  When measured
{\vspace{-0.04cm}}by ${\cal A}_{_{\mbox{\boldmath $\ast$}}}$, a mass loss
over a $\sim$4-hour long period, typically involved, was found from the
ratios ${\cal M}_{\rm fin}/{\cal M}_{\rm beg}$ to be equivalent to a
decrease in the effective dimensions by two to five orders of magnitude.
However, this result overestimates the rate of mass drop{\vspace{-0.04cm}}
because ${\cal A}_{_{\mbox{\boldmath $\ast$}}}$ is, as seen from Table~7,
dominated by the radial component ${\cal A}_1$.  Since much of the mass of
the nucleus of a dwarf sungrazer in this late stage of disintegration is
reduced to expanding clouds of dust, including microscopic particles,
a fraction of the detected acceleration is necessarily contributed
by the Sun's radiation pressure.  Loss effects due to erosion are more
realistically estimated from the transverse and normal components,
which include no contributions from solar radiation pressure.  In that
case the decrease in the effective nuclear size over the 4~hours is
found to amount to $\frac{1}{2}$ to 2$\frac{1}{2}$ orders of magnitude,
a result that suggests the objects' imminent decay.  The heliocentric
distances at which the test dwarf sungrazers were observed ranged from
14.5 to 8\,{\Rsun}.

The presence of nongravitational laws with slopes much steeper than the
square of heliocentric distance even this close to the Sun (which is the
case primarily with C/2007~X3, C/2008~M4, and C/2006~J9) does not necessarily
rule out effects of radiation pressure, because the dust grains continue to
fragment rapidly and the radiation-pressure acceleration varies inversely
as the grain size, except for particles not exceeding in size a small
fraction of a~micron.

We are aware of the limitations that uncertainties in the astrometric data
and short orbital arcs covered by the observations place on the quality of the
orbit determination.  As a result, the orbits in Table~7 should be perceived
with great caution.  Observational errors are likely to be responsible for
the two dynamically meaningless cases of negative ${\cal A}_1$ (C/2007~X3 and
C/2008~M4).  We also readily admit that the orbital sets and nongravitational
parameters of all test sungrazers in Table~7 do not necessarily present unique
solutions.  On the other hand, we notice a high concentration of perihelion
distances between 1.0 and 1.1~{\Rsun} among the nongravitational solutions
in both Table~4 and 7, which is significant.  In the other elements, the
deviations between the two sets of nongravitational solutions do not exceed
a few degrees in $\omega$ and $\Omega$, but only 1$^\circ$ in $i$, and
0.03~day in $t_\pi$.

\section{Summary and Conclusions}
The prime objective of this paper was to understand the discrepancy in
the spatial orientation of apsidal lines between the bright and the dwarf
members of the  Kreutz sungrazing system, as revealed by their catalogued
purely gravitational orbits.  The apsidal lines of seven bright Kreutz
sungrazers, observed from the ground in the years 1843--2011, are nearly
perfectly aligned (to within a small fraction of 1$^\circ$), whereas the
apsidal lines of about 1600 faint, dwarf Kreutz sungrazers, detected only
with the coronagraphs on board the {\it SOHO\/} and {\it STEREO\/} spacecraft
between early 1996 and mid-2010, are distributed along an arc extending
$\sim$25$^\circ$ in perihelion latitude $B_\pi$ but not in perihelion
longitude $L_\pi$, which is statistically invariable and equal to $L_\pi$
of the bright sungrazers.  A corollary of this peculiar effect in a plot
of the orbit inclination against the longitude of the ascending node is
a distribution of the dwarf sungrazers along three parallel curves,
each of which subtends an angle of about 15$^\circ$ with the curve of the
reference apsidal line, populated by the bright sungrazers, and passes
through, respectively, the locations in the plot of comets C/1843~D1,
C/1970~K1, and C/2011~W3.

The differences between the apsidal-line orientation patterns of the dwarf
and the bright Kreutz sungrazers suggest that the dwarf objects failed to
avoid exposure to nongravitational dynamical forces that the bright ones
managed to escape.  From the perturbation theory we find that the dwarf
sungrazers' extension of the apsidal-line distribution in $B_\pi$ is a
product of nongravitational accelerations, directed normal to the orbital
plane.  The broader is the range of these accelerations acting on individual
dwarf sungrazers, the wider is the spread in the objects' angular elements.

We examined several different processes that could potentially generate the
major effects in the latitude $B_\pi$. The only plausible trigger appears to
be the erosion-driven transfer of momentum from outgassing to progres\-sively
fragmenting debris of the original nucleus of the dwarf sungrazer.  This
conceptual model~deems~\mbox{fitting} the motions of dwarf sungrazers by any
purely gravitational orbit inappropriate and emphasizes the need to apply,
instead, a nongravitational orbit whose apsidal line matches the reference
apsidal line as closely as possible.  For each dwarf sungrazer this
condition requires (i)~the incorporation of a nongravitational term with
an unknown parameter into the equations of motion; (ii)~the iteration of
the orbital solution with the nongravitational term until a minimum offset
from the reference apsidal line is reached; and (iii)~the determination
of a final set of orbital elements and the nongravitational parameter as
products of the corrected apsidal-line offset.  A successful implementation
of this corrective procedure requires that the quality of fitting the
astrometric observations (that is, their RMS residual) by the
nongravitational solution be better than, or comparable to, that by the
gravitational solution.

To examine the nature and the magnitudes of the nongravitational forces, we
employed a sample of eight test dwarf sungrazers whose gravitational orbits
were distrib\-uted nearly uniformly along a $\sim$90$^\circ$-long arc in
the longitude of the ascending node, between 305$^\circ$ and 33$^\circ$.
The gravitational solutions left offsets from the reference apsidal line
of up to $\sim$17$^\circ$ (Table~4); to rectify these unaccept\-ably large
deviations, we proceeded in three steps.

In the first step, we applied the standard Style~II formalism by Marsden et
al.\ (1973) that incorporates the nongravitational terms into the equations
of motion.  Because of the strong trends in $B_\pi$, we focused on the normal
component of the erosion-driven acceleration, which we introduced into the
computations iteratively and minimized the offset of each dwarf sungrazer's
orbit from the reference apsidal-line orientation to find the parameter
$A_3$.  We assumed in this step that the radial and transverse components
of the nongravitational acceleration were nil, \mbox{$A_1 \!=\! A_2 \!=\! 0$}.
The resulting parameters $A_3$ for the eight test dwarf sungrazers were on
the orders of 10$^{-4}$ to 10$^{-6}$~AU~day$^{-2}$, at least one order and
up to three orders of magnitude {\it greater\/} than the largest values of
$A_1$ among the non-Kreutz comets catalogued by Marsden \& Williams (2008).
C/2007~X13 was found to be subjected to a {\it nongravitational acceleration
greater than the Sun's gravitational acceleration\/} at the same heliocentric
distance.  The derived parameters approximately correlated with the offsets
from the reference apsidal line left by the gravitational orbital solutions.
These offsets were reduced by the nongravitational solutions substantially,
never to exceed 1$^\circ\!$.4.

Because Marsden et al.'s (1973) formalism employs a momentum-transfer law
based on the assumption that the nongravitational acceleration is driven
by outgassing of water ice, we compared, in our second step, this standard
law with a few novel laws that describe outgassing of other species, more
likely than water ice to match the sublimation conditions at extremely small
distances, about 10~{\Rsun}, from the Sun, where the dwarf Kreutz sungrazers
are typically observed (footnote c to Table~7).  The most powerful among the
tested scenarios was the {\it modified law\/}, with a variable scaling
distance $r_0$, linked to the degree of volatility of outgassing species.
This exercise illustrates a great variety of behavior among the test
sungrazers, from cases suggesting that outgassing is dominated by substances
more refractory than forsterite to those with substances more volatile than
sodium and almost as volatile as water ice.  The magnitudes of the
erosion-driven acceleration did not change substantially from those found
in the first step.  The test objects' apsidal-line offsets offered by the
best fitting nongravitational laws were now further reduced to less
than 0$^\circ\!$.7.

The default condition, \mbox{$A_1 \!=\! A_2 \!=\! 0$}, was relaxed in the
third step, in which only the statistical constraint \mbox{$\Delta L_\pi = 0$}
was still retained.  The expansion from single-component to three-component
nongravitational solutions was computationally much more demanding, even
though an employed relationship between the radial and transverse components
meant an increase by only one parameter.  For each of the test comets we used
a solution based on the law that provided the least offset from the reference
apsidal-line orientation, usually the modified law.  The offsets for all
eight tested sungrazers were now reduced to less than 0$^\circ\!$.2, the
level of intrinsic scatter in the orientation of the reference line of
apsides among the bright sungrazers (Table~1); and, remarkably, in four
cases the offset did not even exceed 0$^\circ\!$.01.

Even though the RMS residual was not a criterion by which we judged the
quality of orbital solutions, the results in Table~7 show that the final
nongravitational sets of orbital elements provide a significantly better
fit than the gravitational solutions (Table~4) in one case and slightly
to moderately better fits in three cases, and that the fits are comparable
in the remaining four cases.

The parameters ${\cal A}_1$ to ${\cal A}_3$ in Table~7 show that at a
distance of 10~{\Rsun} from the Sun, {\it all\/} test dwarf sungrazers were
subjected to nongravitational accelerations of more than 15~percent of
the Sun's gravitational acceleration, or more than two orders of magnitude
higher than implied by the peak outgassing-driven accelerations for the
catalogued comets in nearly parabolic orbits.

There are two more points to emphasize.  The first is that the eight
test sungrazers do by no means represent a random sample.  The condition
that the orbits of the test objects be approximately uniformly distributed
in the longitude of the ascending node required a careful selection of
appropriate candidates.  It was the small number of astrometric observations
made with the C2 coronagraph that caused difficulties.

The second point has to do with the catalogued gravitational orbits of the
dwarf Kreutz sungrazers.  We required in this paper that for each selected
test object the set of orbital elements computed by Marsden be closely
reproduced by our code.  This turned out to be more constraining than first
thought, because in his effort to obtain a sungrazing-like orbit Marsden
sometimes manipulated the values of the elements by assuming a particular
value for the perihelion distance.  The first author has been aware of
Marsden's frustration with this issue for a number of {\it SOHO/STEREO\/}
sungrazers.  He altogether too often obtained a perihelion distance smaller
than the Sun's radius, which --- especially in early years\footnote{Only
after Sekanina's (2002) paper on the dynamical effects of cascading
fragmentation among dwarf Kreutz sungrazers along their entire orbit about the
Sun started Marsden gradually accepting the fact that the perihelion distances
of these objects could be less than 1\,{\Rssun}.} --- he considered unphysical.
On other occasions, the perihelion distance came out to be much too large,
so that, without manipulation, he could not classify the object as a sungrazer
in spite of the signature in the angular elements.  Since Marsden's forcing
a perihelion distance affected the other elements as well, all such cases
curtailed the list of candidates for our test objects.

In this context, the strong concentration of perihelion distances just
beyond 1~{\Rsun}, apparent from Table~7 (as well as from the nongravitational
solutions in Table~4) and already commented on, is notable.  Given a
scatter of 0.8~{\Rsun} in the perihelion distances from the gravitational
orbits of the test sungrazers (Table 4), the apsidal-line constraint should
in no way cause the sharp peak in the distribution.  It therefore appears
that the introduction of the nongravitational terms into the equations of
motion improved substantially the quality of orbit determination, even though
the realistic errors in the elements in Table~7 are greater than the last
decimal places carried.  From comparison of the nongravitational sets of
elements in Table~4 with those in Table~ 7, we estimate that the errors
are on the order of 0.01~day in $t_\pi$, from 0$^\circ\!$.1 to a degree
or so in $\omega$ and $\Omega$, from better than 0$^\circ\!$.1 to nearly
1$^\circ$ in $i$, and from better than 0.01\,{\Rsun} to $\sim$0.1\,{\Rsun}
in $q$; the least well determined orbit appears to be that of C/2007~X3,
which happens to be derived from only 7 astrometric observations.

Application of a basic mass-loss model for the test dwarf sungrazers 
suggests that while observed in the~field of the {\it SOHO\/}'s C2 coronagraph,
along the terminal segment of their orbits, the nuclei fragmented
copiously and their dimensions shrank at a dramatic rate, resulting in
the objects' imminent decay, as illustrated by Schrijver et al.\ (2012)
in the case of C/2011~N3.  Along the radial direction, the nongravitational
effect due to erosion-driven momentum transfer should, in the late stage of
disintegration, have been enhanced, to at least a limited degree, by solar
radiation pressure acting on the microscopic debris of the nucleus.\\[-0.15cm]
\section{What To Do Next?}
The results of our orbital analysis strongly suggest that due to a major
erosion of their mass, the~\mbox{nuclei} of dwarf Kreutz-system comets are
in close proxim\-ity to the Sun subjected to a momentum-transfer effect of
such magnitude that:\ (i)~their motions cannot be fitted by employing the
gravitational law alone, as such orbital solutions lead to grossly
misleading results, and (ii)~the apsidal-line orientations derived from
such solutions require major corrective measures, whose application is
involved and time consuming.  Yet, the gravitational orbits are the only
orbital data currently available for the dwarf Kreutz sungrazers.
A perplexing issue is what to do to rectify this situation?

A skeptic would suggest that the quality of astrometric observations of the
dwarf comets from {\it SOHO/STEREO\/} images is too poor to mount a massive
project in an effort to fix the problem.  In fact, reference to the low accuracy
of the positional data has often been an argument used to question the merit
of these objects' published orbital elements in the first place.  Our results
--- although based on a very limited data sample --- suggest that these
doubts are not necessarily justified and that the problem can in principle
be cured by appropriately accounting for the large erosion-driven
nongravitational forces.

In practical terms, should the orbits of all, or in the least a sizable
fraction, of the dwarf sungrazers of the Kreutz system be reanalyzed from
scratch?  The authors admit that they have neither~re\-sources nor a
motivation for getting involved with such a large-scale but essentially
routine project that should basically follow the algorithm prescribed in this
paper. Accordingly, we limit our comments to acknowledging that the current
state of orbital analysis of the body of these objects is rather depressing,
but we refrain from proposing a workable plan of action beyond merely
recognizing that the task requires highly sophisticated computer-driven
automation.\\

The authors thank A.\ Vitagliano for modifications in his code {\it EXORB7\/}
that he made at our request and thus allowed us to pursue all planned steps
in our investigation.  This research was carried out in part at the Jet
Propulsion Laboratory, California Institute of Technology, under contract
with the National Aeronautics and Space Administration.{\vspace{0.25cm}}
\begin{center}
{\footnotesize REFERENCES}
\end{center}
\vspace*{-0.25cm}
\begin{description}
{\footnotesize
\item[\hspace{-0.3cm}]
Biesecker, D. A., Lamy, P., St.\,Cyr, O. C., et al. 2002, Icarus, 157,
{\linebreak} {\hspace*{-0.6cm}}323
\\[-0.57cm]
\item[\hspace{-0.3cm}]
Brueckner, G. E., Howard, R. A., Koomen, M. J., et al. 1995, Sol.
{\hspace*{-0.6cm}}Phys., 162, 357
\\[-0.57cm]
\item[\hspace{-0.3cm}]
Ciaravella, A., Raymond, J. C., \& Giordano, S.\,2010, ApJ, 713, 69
\\[-0.57cm]
\item[\hspace{-0.3cm}]
Danby, J.\ M.\ A. 1988, Fundamentals of Celestial Mechanics, p.
{\linebreak} {\hspace*{-0.6cm}}323 (2nd ed.; Richmond, VA:
Willmann-Bell Publ., 466pp)
\\[-0.57cm]
\item[\hspace{-0.3cm}]
Gray, W. J. 2013, MPC, 84616 ff.
\\[-0.57cm]
\item[\hspace{-0.3cm}]
Hashimoto, A. 1990, Nature, 347, 53
\\[-0.57cm]
\item[\hspace{-0.3cm}]
Hicks, W. T. 1963, J. Chem. Phys., 38, 1873
\\[-0.57cm]
\item[\hspace{-0.3cm}]
Howard, R. A., Moses, J. D., Vourlidas, A., et al. 2008, Space Sci.
{\hspace*{-0.6cm}}Rev., 136, 67
\\[-0.57cm]
\item[\hspace{-0.3cm}]
Hufnagel, L. 1919, Astron. Nachr., 209, 17
\\[-0.57cm]
\item[\hspace{-0.3cm}]
Kimura, H., Mann, I., Biesecker, D. A., \& Jessberger, E. K. 2002,
{\hspace*{-0.6cm}}Icarus, 159, 529
\\[-0.57cm]
\item[\hspace{-0.3cm}]
Knight, M. M., A'Hearn, M. F., Biesecker, D. A., et al. 2010, AJ,{\linebreak}
{\hspace*{-0.6cm}}139, 926
\\[-0.57cm]
\item[\hspace{-0.3cm}]
Kreutz, H. 1888, Publ. Sternw. Kiel, 3
\\[-0.57cm]
\item[\hspace{-0.3cm}]
Kreutz, H. 1891, Publ. Sternw. Kiel, 6
\\[-0.57cm]
\item[\hspace{-0.3cm}]
Kreutz, H. 1901, Astron. Abh., 1, 1
\\[-0.57cm]
\item[\hspace{-0.3cm}]
Marsden, B. G. 1967, AJ, 72, 1170
\\[-0.57cm]
\item[\hspace{-0.3cm}]
Marsden, B. G. 1970, IAU Circ., 2261
\\[-0.57cm]
\item[\hspace{-0.3cm}]
Marsden, B. G. 2005, ARA\&A, 43, 75
\\[-0.57cm]
\item[\hspace{-0.3cm}]
Marsden, B. G., \& Williams, G. V. 2008, Catalogue of Cometary
{\hspace*{-0.6cm}}Orbits 2008, p.\ 108 (17th ed.; Cambridge, MA:
Smithsonian {\hspace*{-0.6cm}}Astrophysical Observatory, 195pp)
\\[-0.57cm]
\item[\hspace{-0.3cm}]
Marsden, B.\,G., Sekanina, Z., \& Yeomans, D.\,K.\ 1973, AJ,\,78,\,211
\\[-0.57cm]
\item[\hspace{-0.3cm}]
Nakano, S. 2001, Nakano Note 933
\\[-0.57cm]
\item[\hspace{-0.3cm}]
Nakano, S. 2006, Nakano Notes 1295 and 1305
\\[-0.57cm]
\item[\hspace{-0.3cm}]
Nakano, S. 2008, Nakano Note 1618
\\[-0.57cm]
\item[\hspace{-0.3cm}]
Schrijver, C. J., Brown, J. C., Battams, K., et al. 2012, Science,
{\linebreak} {\hspace*{-0.6cm}}335, 324
\\[-0.57cm]
\item[\hspace{-0.3cm}]
Sekanina, Z. 1982, in Comets, ed. L. L. Wilkening (Tucson, AZ:
{\hspace*{-0.6cm}}University of Arizona), 251
\\[-0.57cm]
\item[\hspace{-0.3cm}]
Sekanina, Z. 2000, ApJ, 545, L69
\\[-0.57cm]
\item[\hspace{-0.3cm}]
Sekanina, Z. 2002, ApJ, 566, 577
\\[-0.57cm]
\item[\hspace{-0.3cm}]
Sekanina, Z. 2003, ApJ, 597, 1237
\\[-0.57cm]
\item[\hspace{-0.3cm}]
Sekanina, Z. 2005, Int. Comet Quart., 27, 225
\\[-0.57cm]
\item[\hspace{-0.3cm}]
Sekanina, Z., \& Chodas, P. W. 2007, ApJ, 663, 657
\\[-0.57cm]
\item[\hspace{-0.3cm}]
Sekanina, Z., \& Chodas, P. W. 2008, ApJ, 687, 1415
\\[-0.57cm]
\item[\hspace{-0.3cm}]
Sekanina, Z., \& Chodas, P. W. 2012, ApJ, 757, 127 (33pp)
\\[-0.57cm]
\item[\hspace{-0.3cm}]
Sekanina, Z., \& Kracht, R. 2013, ApJ, 778, 24 (13pp)
\\[-0.57cm]
\item[\hspace{-0.3cm}]
Sekanina, Z., \& Kracht, R. 2014, eprint arXiv:1404.5968
\\[-0.65cm]
%
% \item[\hspace{-0.3cm}]
%
\item[\hspace{-0.3cm}]
Thompson, W. T. 2009, Icarus, 200, 351}
\\[-0.725cm]
%
% \vspace*{-0.6cm}
%
\end{description}
\end{document}